\documentclass[a4paper,11pt]{article}
\usepackage{jheppub} 
\usepackage{jheppub}
\usepackage[utf8]{inputenc}
\usepackage{CJK}
\usepackage{epstopdf}
\usepackage{enumitem}
\usepackage{tcolorbox}
\usepackage{tikz}
\usetikzlibrary{positioning}
\usepackage{jheppub}
\usepackage{mathrsfs}
\usepackage{psfrag}
\usepackage{color}
\usepackage{hyperref}
\usepackage{slashed}
\usepackage{feynmp-auto}
\usepackage{simplewick}
\usepackage{cancel}

\usepackage{todonotes}

\usepackage{amsmath}
\usepackage{amsfonts}
\usepackage{graphicx}
\usepackage{amssymb}
\usepackage{xcolor}

\usepackage{mathtools}

\usepackage{amsmath,bbm,array,amsfonts,graphicx,wrapfig,arydshln,lscape,float,multirow,longtable,rotating,makecell}
\usepackage{url}

\def\Label#1{\label{#1}%
	\smash{\hbox to0pt{\raise1ex\hbox{\tiny[#1]}\hss}}}
\def\Label#1{\label{#1}}
\renewcommand{\eqref}[1]{eq.~(\ref{#1})}
\newcommand{\p}{\partial}

\def\abs#1{\left| #1\right|}
\def\braket#1{\langle #1 \rangle}
\def\sbraket#1{[ #1 ]}
\newcommand{\bra}[1]{\langle{#1}|}
\newcommand{\ket}[1]{|{#1}\rangle}

\newcommand{\sbra}[1]{ [{#1} |}
\newcommand{\sket}[1]{ | {#1} ]}

\newcommand{\ep}{\epsilon}

\newcommand{\dd}{\mathrm{d}}

\newcommand{\ap}{\alpha'}

\title{\boldmath Higher-Point Gauge-Theory Couplings of Massive Spin-2 States in 4-Dimensional String Theories}

\author{Chen Huang (黄晨)}
\affiliation{Uppsala University,\\75237 Uppsala, Sweden}
\affiliation{Institute of Theoretical Physics, Chinese Academy of Sciences, Beijing 100190, China}

\emailAdd{chen.huang.8145@student.uu.se}

\abstract{We explicitly compute the (NS) sector conventional type-I superstring tree-level amplitudes at five points after compactifying to 4-D, express the QFT building block in the helicity basis, and give several attempts towards arbitrary $n$ points. More specifically, we consider the interaction of one first excited level and otherwise massless states of conventional type-I superstrings, where the four-dimensional states can, for instance, be realized via D$3$ branes. We construct the amplitude by using the Berends-Giele currents. From the recursion of Berends-Giele currents, we can generate the higher point amplitude. We also apply the BCFW recursion with massive external legs shifted and get the amplitude for arbitrary $n$ points.}

\begin{document}
\begin{CJK*}{UTF8}{}
\CJKfamily{gbsn}
\maketitle
\end{CJK*}
\flushbottom

\section{Introduction}
The study of scattering amplitudes is one of the historical origins of string theory \cite{Veneziano:1968yb}. For example, Veneziano amplitude, a candidate amplitude for hadron scattering, is often referred to as the first equation of string theory. Also, some important features of string theory and quantum field theory are hidden in the amplitudes. The computation of string amplitudes is closely related to the correlation functions of vertex operators on the so-called world sheet. Technically, this is a two-dimensional conformal field theory (CFT) on the world sheet. The CFT approach leads to many interesting properties, one of them is the famous KLT relation \cite{Kawai:1985xq}: For closed strings, the genus-zero correlators and in fact, even their integrals over the vertex points can be factorized into left movers and right movers, also called holomorphic and antiholomorphic building blocks. KLT relation indicates the tree-level double-copy relation, especially between perturbative gravity and gauge theories. 
\par In recent decades, numerous studies have been conducted on this topic, see for instance \cite{Kawai:1985xq, Bern:2008qj, Bern:2019prr}, the KLT and double copy relation at tree level have already become one of the essential features of string amplitudes. There are also attempts toward loop-level generalizations \cite{DHoker:1989cxq, Bern:2010ue, He:2016mzd, Stieberger:2022lss}. In the past ten years, the discoveries of additional double-copy structures indicated that, if all external states are massless, the tree-level coupling of the type-I superstring \cite{Mafra:2011nv, Broedel:2013tta} and the open bosonic string \cite{Huang:2016tag, Azevedo:2018dgo} can be factorized into the scalar integrals on disk (the disk integral is also known as Z-theory amplitudes) and quantum field theory (QFT) building blocks. Some recent papers even generalized this relation to the coupling with $1$ external mass-level-$1$ state and found the QFT building block of this coupling \cite{Guillen:2021mwp, Kashyap:2023cdi, HuangShruti}.
\par When analyzing the QFT amplitude, the complexity and computational difficulty of the traditional Feynman diagrammatic approach increase rapidly as the number of external particles grows \cite{Parke:1986gb}. In contrast, the development of modern amplitude methods in the past few decades has directed an alternative route to arrive at otherwise intractable results, both classical and quantum \cite{Elvang:2015rqa, Parke:1986gb}, and various recursion relations. Apart from the famous on-shell BCFW recursion \cite{Britto:2005fq}, there is also a semi-on-shell, recursion relation based on the Berend-Giele (B-G) currents \cite{Berends:1987me, Berends:1988zp, Berends:1988zn}. There are also applications to amplitudes of various theories at tree and loop level \cite{Tao:2023wls, Chen:2023bji, Tao:2023yxy, Tao:2022nqc}. 
\par Since the KLT and the additional double-copy relation are key features of string amplitudes and KLT can work for arbitrary excited states of strings, it is really important to understand how the additional double-copy structure behaves when excited states are involved. Since all the information on polarization and momentum is inside the QFT building block, it is important to learn the dynamical structure of the QFT building block, especially in 4-D after compactification or due to the D branes. One of the most powerful tools in researching 4-D QFT amplitudes is the so-called spinor helicity formalism. Thus, our aim in this paper is to produce the spinor helicity form of the QFT building block for conventional type-I string amplitude, we first reviewed the basic idea of spinor helicity, BCFW recursion, dimension-agnostic Berends-Giele currents, and B-G recursion. Then generalized the result involving one of the universal Regge excitation states and up to three massless gluons \cite{Feng:2010yx} to one excitation state coupling with arbitrary $n-1$ massless gluons, which result in a very nice $n$-point amplitude formula \eqref{formula} for a specific configuration of external gluon helicities. This helicity configuration resembles the so-called maximally helicity violation (MHV) in pure Yang-Mills. We analytically proved the $5$-point case and gave a numerical check for the $6$-point case. We also applied BCFW to the QFT building blocks of the conventional type-I superstring amplitude. The final result is in agreement with \eqref{formula}.
\section{Notation and convention}
	\subsection{Spinor helicity in 4 dimensions}
	\par Spinor helicity formalism \cite{Elvang:2015rqa} is another way\footnote{With respect to the covariant way, which preserves the locality and Lorentz invariance in each Feynman diagram.} of expressing the QFT amplitude. It has extremely simplified the calculation of the Scattering amplitudes in $4$-dimension \cite{Parke:1986gb}. There is also spinor helicity formalism in other dimensions, for instance, $3,\enspace 6,$ and $10$ dimension \cite{Caron-Huot:2010nes}, but our discussion is specified to $4$-dimension, thus we only introduce the $4$-dimensional spinor helicity formalism in this paper.
	\par In this subsection, we will introduce the conventions for the spinor helicity formalism:
	\par We will mostly keep those in Elvang and Huang's Scattering amplitudes textbook \cite{Elvang:2015rqa}. The metric is chosen to be the "mostly-plus" metric, $\eta_{\mu\nu}=\mathrm{diag}(-1,+1,+1,+1)$. We define the $\sigma$ matrices\footnote{We will omit the spinor indices $(\alpha,\beta,\dots)$ later on when it leads to no misunderstanding. When we come to the massive spinor part, we use $(a,b,\dots)$ as the little group indices. We will come to the details in that section.} as:
\begin{equation}
    (\sigma^{\mu})_{\alpha\dot{\beta}}=(1,\sigma^{i})_{\alpha\dot{\beta}},\qquad(\bar{\sigma}^{\mu})^{\dot{\alpha}\beta}=(1,-\sigma^{i})^{\dot{\alpha}\beta},
\end{equation}
where $\sigma^{i}$, $i=1,2,3$ are Pauli matrices:
\begin{equation}
		\sigma^{1}=	
		\begin{pmatrix}
			0 & 1 \\
			1 & 0 \\
		\end{pmatrix}
		\qquad
		\sigma^{2}=	
		\begin{pmatrix}
			0 & -i \\
			i & 0 \\
		\end{pmatrix}
		\qquad
		\sigma^{3}=	
		\begin{pmatrix}
			1 & 0 \\
			0 & -1 \\
		\end{pmatrix}.
	\end{equation}
	The two sets of spinor indices are raised and lowered individually using the $SU(2)$ invariant tensor, also known as the Levi-Civita tensor:
	\begin{equation}
		\varepsilon^{\alpha \beta}=\varepsilon^{\dot{\alpha}\dot{\beta}}=
		\begin{pmatrix}
			0 & 1 \\
			-1 & 0 \\
		\end{pmatrix}
		=-\varepsilon_{\alpha \beta}=-\varepsilon_{\dot{\alpha}\dot{\beta}},
	\end{equation}
	obey $\varepsilon_{\alpha \beta}\varepsilon^{\beta \gamma}=\delta_{\alpha}^{\enspace \gamma}$
	\par We contract $(\sigma^{\mu})_{\alpha\dot{\alpha}}$ with $k^{\mu}$ to get the $2\times2$ matrix $k_{\alpha\dot{\alpha}}=k_{\mu}\sigma^{\mu}_{\alpha\dot{\alpha}}$ also notice $\mathrm{det}(k_{\alpha\dot{\alpha}})=m^{2}$, which leads to the obvious difference between the massive external states and the massless external states, we will come to this later.
	\subsubsection{Spinor helicity for massless particles}
	\par For massless particles, we have $\mathrm{det}(k_{\alpha\dot{\alpha}})=0$ and thus the matrix $k_{\alpha\dot{\alpha}}$ is of rank 1. We can then write it as the direct product form:
	\begin{equation}
		k_{\alpha\dot{\alpha}}=-\lambda_{\alpha}\tilde{\lambda}_{\dot{\alpha}},
	\end{equation}
	we also write $\lambda_{\alpha}$ and $\tilde{\lambda}_{\dot{\alpha}}$ as $\sket{k}_{\alpha}$ and $\bra{k}_{\dot{\alpha}}$. So we can also write the momentum in a matrix form:
	\begin{equation}\label{spinorhelicity}
		k_{\alpha\dot{\alpha}}=-\sket{k}_{\alpha}\bra{k}_{\dot{\alpha}}.
	\end{equation}
	Consider the Dirac equation in the massless case. The Dirac equation would decouple into the Weyl equation without mass. Thus, when $m=0$, we have:
	\begin{equation}
		\slashed{k}v_{\pm}(k)=0,\qquad \bar{u}_{\pm}(k)\slashed{k}=0,
	\end{equation} 
	where $v_{\pm}(p)$ and $\bar{u}_{\pm}(p)$ are wave functions associated with an outgoing anti-fermion and fermions. The wave functions are related as $u_{\pm}=v_{\mp}$ and $\bar{v}_{\pm}=\bar{u}_{\mp}$
	\par We can now write the two independent solutions of the Dirac equation as:
	\begin{equation}
		v_{+}(k)=
		\begin{pmatrix}
			&\sket{k}_{\alpha}\\
			&0
		\end{pmatrix}
		\qquad
		v_{-}(k)=
		\begin{pmatrix}
			&0\\
			&\ket{k}^{\dot{\alpha}}
		\end{pmatrix},
	\end{equation}
	and
	\begin{equation}
		\bar{u}_{-}(k)=
		\begin{pmatrix}
			0, \enspace\bra{k}_{\dot{\alpha}}
		\end{pmatrix}
		\qquad
		\bar{u}_{+}(k)=
		\begin{pmatrix}
			\sbra{k}^{\alpha},\enspace 0
		\end{pmatrix}.
	\end{equation}
	The angle and square spinors are 2-component commuting spinors. After defining the spinor, we can write the massless Weyl equation:
	\begin{equation}
		k^{\dot{\alpha}\beta}\sket{k}_{\beta}=0,\quad k_{\alpha\dot{\beta}}\ket{k}^{\dot{\beta}}=0,\quad \sbra{k}^{\beta}k_{\beta\dot{\alpha}}=0,\quad 
		\bra{k}_{\dot{\beta}}k^{\dot{\beta}\alpha}=0.
	\end{equation}
we have two sets of important identities. First, the Fierz identities:
\begin{equation}\label{Fierz}
\begin{aligned}
    (\sigma^{\mu})_{\alpha\dot{\alpha}}(\sigma_{\mu})_{\beta\dot{\beta}}&=-2\varepsilon_{\alpha\beta}\varepsilon_{\dot{\alpha}\dot{\beta}},
    \\
    (\sigma^{\mu})_{\alpha\dot{\alpha}}(\bar{\sigma}_{\mu})^{\beta\dot{\beta}}&=-2\delta_{\alpha}^{\enspace \beta}\delta_{\dot{\alpha}}^{\enspace\dot{\beta}},
    \\
    \sbra{a}\sigma^{\mu}\ket{b} \sbra{c}\sigma_{\mu}\ket{d}&=2\sbraket{ac}\braket{bd},
    \\
    \sbra{a}\sigma^{\mu}\ket{b} \bra{c}\bar{\sigma}_{\mu}\sket{d}&=2\braket{bc}\sbraket{ad},
\end{aligned}
\end{equation}
and second, the Schuten identity:
\begin{equation}
    \ket{i}\braket{jk}+\ket{j}\braket{ki} +\ket{k}\braket{ij}=0
\end{equation}
	\subsubsection{Spinor helicity for massive particles}
	We now consider the massive spinor helicity formalism \cite{Novaes:1991ft,Spehler:1991yw}. The spinor helicity formalism of massive particles have no essential difference from the one of massless particles. We can regard the massive momentum as the linear combination of two massless momenta:
	\begin{equation}
		k^{\mu}_{i}=a^{\mu}_{i}+b^{\mu}_{i},
	\end{equation}
	where $a^{\mu}_{i},b^{\mu}_{i}$ have the same spatial direction as $k^{\mu}_{i}$, which are along z axis. We can write down the components of $k, a,b$ now:
	\begin{equation}
		k^{\mu}_{i}=\left(k^{0}_{i},0,0,k^{3}_{i}\right)\quad a^{\mu}_{i}=\left(\frac{k^{0}_{i}+k^{3}_{i}}{2},0,0,\frac{k^{0}_{i}+k^{3}_{i}}{2}\right)\quad b^{\mu}_{i}=\left(\frac{k^{0}_{i}-k^{3}_{i}}{2},0,0,-\frac{k^{0}_{i}-k^{3}_{i}}{2}\right).
	\end{equation} 
	Or, more generally, (both direction and the length), we can parameterize the massive momentum $k^{\mu}_{i}$:
	\begin{equation}\label{decomposition_of_p}
		k^{\mu}_{i}=a^{\mu}_{i}-\frac{m_{i}^{2}}{2a_{i}\cdot b_{i}}b^{\mu}_{i},
	\end{equation}
	we can contract both side of this equation with $\sigma^{\mu}$, and decompose $a^{\mu}_{i}$ and $b^{\mu}_{i}$ on the right hand side into spinor helicity form using \eqref{spinorhelicity}, we have:
	\begin{equation}
		k_{\alpha\dot{\alpha}}=\sket{a}_{\alpha}\bra{a}_{\dot{\alpha}}+\sket{b}_{\alpha}\bra{b}_{\dot{\alpha}},
	\end{equation}
	this is simply a rank 2 matrix $k_{\alpha\dot{\alpha}}$. Which can now be written as:
	\begin{equation}
		k_{\alpha\dot{\alpha}}=\lambda^{\enspace a}_{\alpha}\tilde{\lambda}_{\dot{\alpha}a},
	\end{equation}
	where $a=1,2$ corresponds to $\sket{a},\enspace\ket{a}$ and $\sket{b},\enspace\ket{b}$ separately. We now rewrite $\lambda$ and $\tilde{\lambda}$ as matrix, regard $\alpha$ and $a$ as the matrix index. We then have:
	\begin{equation}
		k^{2}=-m^{2}\rightarrow\mathrm{det}\lambda\times\mathrm{det}\tilde{\lambda}=-m^{2},
	\end{equation}
where we set\footnote{This is a trivial convention, $\mathrm{det}\lambda$ and $\mathrm{det}\tilde{\lambda}$ are not necessarily equal to each other, there could be a phase factor that makes $\mathrm{det}\lambda$ and $\mathrm{det}\tilde{\lambda}$ both different from $m$ but preserve the constraint $\mathrm{det}\lambda\times\mathrm{det}\tilde{\lambda}=m^{2}$, let's take the trivial one as an example, we will come to this very soon.} $\mathrm{det}\lambda=\mathrm{det}\tilde{\lambda}=m$
	We can also raise or lower the indices $a,b$ by using $\epsilon^{ab}$ and $\epsilon_{ab}$ so that we can write:
	\begin{equation}
		k_{\alpha\dot{\alpha}}=\lambda^{\enspace a}_{\alpha}\tilde{\lambda}_{\dot{\alpha}}^{\enspace b}\epsilon_{ab}.
	\end{equation}
	Also, notice that we have the Dirac equation:
	\begin{equation}\label{Dirac1}
		k_{\alpha\dot{\alpha}}\tilde{\lambda}^{\dot{\alpha} a}=\lambda^{\enspace b}_{\alpha}\tilde{\lambda}_{\dot{\alpha}b}\tilde{\lambda}^{\dot{\alpha} a}=\lambda^{\enspace b}_{\alpha}\tilde{\lambda}_{\dot{\alpha}b}\epsilon^{\dot{\alpha} \dot{\beta}}\tilde{\lambda}_{\dot{\beta} c}\epsilon^{ca}=\lambda^{\enspace b}_{\alpha}\mathrm{det}(\tilde{\lambda})\delta_{b}^{\enspace a}=m\lambda^{\enspace a}_{\alpha}.
	\end{equation}
	Similarly, we also have:
	\begin{equation}\label{Dirac2}
		k_{\alpha\dot{\alpha}}{\lambda}^{\alpha a}=-m\tilde\lambda^{\enspace a}_{\dot{\alpha}}.
	\end{equation}
	These two equations are equivalent to Dirac equation \cite{Ochirov:2018uyq}.
	
	\par By using the \cite{Chiodaroli:2021eug} decomposition of massive momentum, we can also write down the component of massive spinor\footnote{This convention is more non-trivial than the one mentioned before. We now still have $\braket{ab}\sbraket{ab}=m^{2}$, but $\braket{ab}=\sbraket{ab}=m$ relationship no longer exist. }:
	\begin{equation}
		\ket{k^{a}}=
		\begin{pmatrix}
			\ket{b}\frac{m}{\braket{ab}}
			\\
			\ket{a}
		\end{pmatrix}
		\qquad
		\sket{k^{a}}=
		\begin{pmatrix}
			\sket{a}
			\\
			\sket{b}\frac{m}{\sbraket{ab}}
		\end{pmatrix}.
	\end{equation}
	In other references, people also define $\bra{\boldsymbol{k}}$ as the massive spinor. In our notation, it is simply $z_{a}$ that is used to absorb the $su(2)$ index.
	\begin{equation}
		\bra{\boldsymbol{k}}=\bra{k^{a}}z_{a}.
	\end{equation}
	We have a constrain on $z_{a}$, $\bar{z}_{b}$ and the antisymmetric tensor $\epsilon^{ab}$:
	\begin{equation}
			z_{a}\varepsilon^{ab}\bar{z}_{b}=-1.
	\end{equation} 
    \par We can construct the polarization tensor\footnote{We will shortly see, this is simply the single particle B-G current reduced to a specific $4$-dimensional choice of polarization.} for the spin-2 massive particles as follows\footnote{The $(k_{i},-2)$ in $\Phi^{\mu\nu}(k_{i},-2)$ means the polarization is a function of the momentum, for spin choice $-2$, we will omit this bracket when it leads to no confusion.}:
    \begin{equation}\label{pov}
        \Phi_{i}^{\mu\nu}(k_{i},-2)=\frac{1}{2m^{2}}\sbra{a}\bar{\sigma}^{\mu}\ket{b}\sbra{a}\bar{\sigma}^{\nu}\ket{b},
    \end{equation}
	where $\Phi^{\mu\nu}$ is a traceless symmetric tensor. The momentum $k$ is decomposed as $k^{\mu}=a^{\mu}+b^{\mu}$. For a massive spin $j$ particle, there exist $2j+1$ spin degrees of freedom, the spin quantization axis is chosen as the direction of $a$ in the rest frame. Each spin choice corresponds to a state, we express all of them by $\ket{m,j}$, where $m=-j,-j+1,\dots,j-1,j$, the $2j+1$ choices of $m$ exactly correspond to the $2j+1$ degrees of freedom we found for spin $j$ particle. 
	\par We can relate the $2j+1$ states by acting the raising and lowering operator on one state and raise or lower $m$:
	\begin{equation}
	    \begin{aligned}
	        J^{+1}\ket{m-1,j}&=\sqrt{\frac{(j+m)(j-m+1)}{2}}\ket{m,j},
	        \\
	        J^{-1}\ket{m,j}&=\sqrt{\frac{(j+m)(j-m+1)}{2}}\ket{m-1,j},
	    \end{aligned}
	\end{equation}
	where the boundary requirements are $J^{+1}\ket{j,j}=0,\enspace J^{-1}\ket{-j,j}=0$.
	\par Similar to the operator acting on states, we can define a set of raising and lowering operators acting on the polarization tensor:
	\begin{equation}\label{rloperator}
	    \begin{aligned}
	        O^{+1}\Phi^{\mu_{1}\mu_{2}\cdots\mu_{j}}(k,m-1)=&\frac{1}{\sqrt{2}}(+\ket{a}\frac{\p}{\p\ket{b}}-\sbra{b}\frac{\p}{\p\sbra{a}})\Phi^{\mu_{1}\mu_{2}\cdots\mu_{j}}(k,m-1)
	        \\
	        =&N(m,j)\Phi^{\mu_{1}\mu_{2}\cdots\mu_{j}}(k,m),
	        \\
	        O^{-1}\Phi^{\mu_{1}\mu_{2}\cdots\mu_{j}}(k,m)=&\frac{1}{\sqrt{2}}(-\ket{a}\frac{\p}{\p\ket{b}}+\sbra{b}\frac{\p}{\p\sbra{a}})\Phi^{\mu_{1}\mu_{2}\cdots\mu_{j}}(k,m-1)
	        \\
	        =&N(m,j)\Phi^{\mu_{1}\mu_{2}\cdots\mu_{j}}(k,m-1),
	    \end{aligned}
	\end{equation}
	where we used $N(m,j)$ as a shorthand of$\sqrt{\frac{(j+m)(j-m+1)}{2}}$, and get the polarization tensor corresponding to $\ket{-1,2},\ket{0,2}\ket{+1,2}\ket{+2,2}$ by acting raising and lowering operator on $\ket{-2,2}$ state\footnote{Here we only write two examples.}:
	\begin{equation}\label{2.25}
	    \begin{aligned}
	        \Phi^{\mu\nu}(k,-1)=&\frac{1}{\sqrt{2}}O^{+1}\Phi^{\mu\nu}(k,-2)
	        \\
	        =&\frac{1}{4m^{2}}\left[\left(\sbra{a}\bar{\sigma}^{\mu}\ket{a}-\sbra{b}\bar{\sigma}^{\mu}\ket{b}\right)\sbra{a}\bar{\sigma}^{\nu}\ket{b}+\sbra{a}\bar{\sigma}^{\mu}\ket{b}\left(\sbra{a}\bar{\sigma}^{\nu}\ket{a}-\sbra{b}\bar{\sigma}^{\nu}\ket{b}\right)\right],
	        \\
	        \Phi^{\mu\nu}(k,0)=&\frac{1}{\sqrt{3}}O^{+1}\Phi^{\mu\nu}(k,-1)
	        \\
	        =&\frac{1}{2m^{2}\sqrt{6}}\Big[\left(\sbra{a}\bar{\sigma}^{\mu}\ket{a}-\sbra{b}\bar{\sigma}^{\mu}\ket{b}\right)\left(\sbra{a}\bar{\sigma}^{\nu}\ket{a}-\sbra{b}\bar{\sigma}^{\nu}\ket{b}\right)
	        \\
	        &-\sbra{a}\bar{\sigma}^{\mu}\ket{b}\sbra{a}\bar{\sigma}^{\nu}\ket{b}-\sbra{b}\bar{\sigma}^{\mu}\ket{a}\sbra{b}\bar{\sigma}^{\nu}\ket{a}\Big].
	    \end{aligned}
	\end{equation}
	\subsection{Dimension-agnostic Berends-Giele recursions}\label{BG}
	\par From now on, we need to deal with the multi-particle Berends-Giele currents and field strength. We use Latin letters $P, Q, X, Y,\dots$ to denote different sets of particles.
	\par The so-called Berends-Giele (B-G for short) recursions \cite{Berends:1987me,Dixon:1996wi} is an effective approach to determining the tensor structure of arbitrary D-dimensional tree amplitudes in pure Yang-Mills theory, introduced by Berends and Giele in 1987 \cite{Berends:1987me}. The idea of B-G recursions is to recursively combine all color-ordered Feynman diagrams with multiple external on-shell legs and one single off-shell leg using the B-G currents $\epsilon^{\mu}_{12,\dots,p}$. They can be regarded as functions of dynamical variables such as polarization vectors $\epsilon^{\mu}_{i}$ and null momentum vectors $k^{\mu}_{i}$ of the external particles $i=1,2,\dots,p$ constrained by the following on-shell conditions:
	\begin{equation}
		\epsilon_{i}\cdot k_{i}=k_{i}\cdot k_{i}=0,
	\end{equation}
	where $i=1,2,\dots,p$ refer to external-state labels, the Lorentz-indices are denoted by Greek letters $\mu,\nu,\dots=0,1,\dots, D-1$.
	\par The B-G recursion of the Yang-Mills amplitude is done via the recursion\footnote{The boundary constraint is that the single particle current equals to the single particle polarization} of B-G current \cite{Berends:1987me}:
	\begin{equation}
	s_{P}\epsilon^{\mu}_{P}=\sum_{XY=P}[\epsilon_{X},\epsilon_{Y}]^{\mu}+\sum_{XYZ=P}\{\epsilon_{X},\epsilon_{Y},\epsilon_{Z}\}^{\mu},
	\end{equation}
	where capital Latin letters $P, Q, X, Y,\dots$ are multiple particle labels, also known as the rank of the B-G current. The length of, for example, $P=12\dots p$ is denoted by $\abs{P}=p$. $[\epsilon_{X},\epsilon_{Y}]^{\mu}$ and $\{\epsilon_{X},\epsilon_{Y},\epsilon_{Z}\}^{\mu}$ are defined as:
	\begin{equation}
		[\epsilon_{X},\epsilon_{Y}]^{\mu}=(k_{Y}\cdot \epsilon_{X})\epsilon^{\mu}_{Y}-(k_{X}\cdot \epsilon_{Y})\epsilon^{\mu}_{X}+\frac{1}{2}(k^{\mu}_{X}-k^{\mu}_{Y})(\epsilon_{X}\cdot \epsilon_{Y}),
	\end{equation}
	\begin{equation}
		\{\epsilon_{X},\epsilon_{Y},\epsilon_{Z}\}^{\mu}=(\epsilon_{X}\cdot \epsilon_{Z})\epsilon^{\mu}_{Y}-\frac{1}{2}(\epsilon_{X}\cdot \epsilon_{Y})\epsilon^{\mu}_{Z}-\frac{1}{2}(\epsilon_{Y}\cdot \epsilon_{Z})\epsilon^{\mu}_{X},
	\end{equation}
	and the Mandelstam variable with multiple particle indices $s_{P}$ is defined as:
	\begin{equation}
		s_{P}=\frac{1}{2}k^{2}_{P},
	\end{equation}
	where the multiple particle momentum $k^{\mu}_{P=12\dots p}=k^{\mu}_{1}+k^{\mu}_{2}+\dots+k^{\mu}_{p}$ 
	\par To review the recursion of B-G current, we need to define the division of multi-particle labels $P=12\dots p$. The summation over $XY=P$ means dividing $P$ into non empty sets $X=12\dots j$, $Y=j+1\dots p$. Where $X$, $Y$ nonempty indicates that $1\leq j\leq p-1$, thus this summation has $\abs{P}-1=p-1$ terms. The same discussion can be applied to the summation over $XYZ=P$. We can also define the field strength $F$:
	\begin{equation}\label{2.31}
		F^{\mu\nu}_{P}=k^{\mu}_{P}\epsilon^{\nu}_{P}-k^{\nu}_{P}\epsilon^{\mu}_{P}-\sum_{XY=P}(\epsilon^{\mu}_{X}\epsilon^{\nu}_{Y}-\epsilon^{\nu}_{X}\epsilon^{\mu}_{Y}),
	\end{equation}
	and get the simpler form of the B-G current:
	\begin{equation}\label{2.32}
		\epsilon^{\mu}_{P}=\frac{1}{2s_{P}}\sum_{XY=P}[(k_{Y}\cdot \epsilon_{X})\epsilon^{\mu}_{Y}+\epsilon^{\nu}_{X}F^{\nu\mu}_{Y}-(X\leftrightarrow Y)].
	\end{equation}
	\par The color-ordered on-shell amplitudes at $n=p+1$ points are recovered by taking the off-shell leg in the rank-$p$ B-G current $\epsilon^{\mu}_{P}$ on shell. This is done by: 
	\begin{itemize}
		\item Contracting with the polarization vector of particle $n$: $\epsilon^{\mu}_{n}$, which is also a B-G current.
		\item Removing the propagator $s^{-1}_{12\dots p}$ in the $p$-particle channel of $\epsilon^{\mu}_{P}$ which would diverge when taken particle $n$ on shell.
	\end{itemize}
	\par Thus, we have:
	\begin{equation}
		\mathcal{A}(1,2,\dots ,n-1,n)=s_{12\dots n-1}\epsilon^{\mu}_{12\dots n-1}\epsilon^{\nu}_{n}\eta_{\mu\nu}.
	\end{equation}
	
	\subsection{Twisted heterotic string and conventional type-I superstrings}
	\par The twisted heterotic string \cite{Hohm:2013jaa, Huang:2016bdd, LipinskiJusinskas:2019cej, Koh:1987hm} is a special kind of string that satisfies the twisted level-matching condition. One of the most important feature of twisted heterotic string is that the spectrum is finite as opposed to the infinite excited states for type II-A, II-B or type I superstring theory. The physical vertex operators represent the following three multiplets of 10D ${\cal N}=1$ supersymmetry:
	\begin{itemize}
		\item a gauge multiplet involving gluon ($A$) and gluino (${\cal X})$,
		\begin{equation} \label{gauge-vertex}
			{\cal V}^a_{A}  = \bar V^a_{\bar J} \otimes V_{\ep} \, e^{i k \cdot X}  \qquad  {\cal V}^a_{{\cal X}}  = \bar V^a_{\bar J} \otimes V_{\chi} \, e^{i k \cdot X},
		\end{equation}
		\item a supergravity multiplet involving graviton, $B$-field and dilaton ($ \bar V_{\bar \epsilon} \otimes V_{\ep}$) as well as gravitino and dilatino ($ \bar V_{\bar \epsilon} \otimes V_{\chi}$),
		\item a massive multiplet with $k^2 = - \frac{4}{\ap}$ comprising a spin-$2$ field $\Phi_{\mu\nu}$, a 3-form $E_{\mu\nu\rho}$ and a spin-$\tfrac{3}{2}$ field $\Psi^{\alpha}_{\mu}$,
		\begin{equation}\label{massive-vertex}
			{\cal V}_{\{\Phi,E,\Psi\}}  = \bar V_T \otimes V_{\{\Phi,e,\psi\}} \, e^{i k \cdot X}.
		\end{equation}
	\end{itemize}
	where the massive states can be viewed as a double copy of a tachyon, $\bar V_T=1$, with the first mass level of the open superstring \cite{Koh:1987hm}. The Lagrangian description of the amplitude with $1$ external massive states and otherwise gauge multiplets is given in \cite{Guillen:2021mwp}. We only discuss the tree-level couplings of the conventional type-I superstring with only one massive multiplet $\underline{n}$:
	\begin{equation}\label{QFTbuildingblock}
		\mathcal{M}_{type-I}(1,2,\dots,n-1,\underline{n})=\sum_{\rho\in S_{n-3}}F^{\rho}(\mathbf{s}_{n})\times \mathcal{A}(1,\rho(2,\dots,n-2),n-1|\underline{n})\left|_{\ap\rightarrow 4\ap},\right.
	\end{equation}
	where $\mathcal{A}(\dots)$ is a rational function of the external momenta as usual for QFT amplitudes. Hence, $\mathcal{A}(\dots)$ in \eqref{QFTbuildingblock} will be later on referred to as "QFT building blocks". Moreover, the disk integral $F^{\rho}(\mathbf{s}_{n})$ is given by:
	\begin{equation}
		\begin{aligned}
			F^{\rho}(\mathbf{s}_{n})=(2\ap)^{n-3}\int_{\Omega}&\dd z_{2}\dd z_{3}\dots\dd z_{n-2}\prod_{1\leq i< j}^{n-1}|z_{ij}|^{2\ap s_{ij}}
			\\
			&\times\rho\left\{\frac{s_{21}}{z_{21}}\left(\frac{s_{31}}{z_{31}}+\frac{s_{32}}{z_{32}}\right)\cdots\left(\frac{s_{n-2,1}}{z_{n-2,1}}+\cdots+\frac{s_{n-2.n-3}}{z_{n-2,n-3}}\right)\right\},
		\end{aligned}
	\end{equation}
	where $z_{ij}=z_{i}-z_{j}$, $\Omega$ stands for the integration area. Here, it is $0<z_{2}<z_{3}<\dots<z_{n-2}<1$, because we have fixed $(z_{1},z_{n-1},z_{n})\rightarrow(0,1,\infty)$. 
	\par We identify multi-particle polarizations with B-G currents \cite{Berends:1987me} mentioned in section \ref{BG}, and write down the recursion rules of B-G currents. Similar with \eqref{2.31} and \eqref{2.32}, we have:
	\begin{equation}\label{currentsfieldstrengthandgauge}
			\Phi_{P}^{\mu\nu} =  \sqrt{\ap} \sum_{P=QR}  F_{Q \, \rho}^{\mu}F_{R}^{\rho\nu}  +{\rm cyc}_P,
	\end{equation}
where the $F^{\mu\nu}_{P}$ and the $\epsilon^{\mu}_{P}$ has the same form as Yang-Mills B-G current, following the discussion in \cite{Guillen:2021mwp}. ${\rm cyc}_{P}$ instructs to add cyclic permutations in $P=1,2,\dots,p$. For example, $A_{P}=B_{P}+{\rm cyc}_P$ where $P={1,2,3,4}$ means:
	\begin{equation}
	    A_{P}=B_{P}+{\rm cyc}_P=B_{1,2,3,4}+B_{2,3,4,1}+B_{3,4,1,2}+B_{4,1,2,3}.
	\end{equation}
The amplitude can be constructed from B-G currents:
	\begin{equation}\label{amplitude}
		{\cal A}(1,2,\ldots,n{-}1|\underline{n})  = (\Phi_{12 \ldots n-1})^{\mu\rho}(\Phi_{n})_{\mu\rho}.
	\end{equation}
	\par For the lowest or highest spin state of the only massive particle, the single particle B-G current\footnote{Now can be regarded as polarization tensor.} $\Phi^{\mu\nu}$ is simply a direct product of the polarization vector of the lowest or highest spin state of spin-$1$ massive particle: $\ep^{\mu}\ep^{\nu}$. Our aim is to prove this $n$ point amplitude equals to \eqref{formula}.
	\subsection{Amplitudes compactified to 4-D}\label{4dim}
	We impose the external polarization and momentum to lie in $4$ dimensional Minkowski space, and convert the amplitude into spinor helicity form.
	\par Some useful expressions are stated below:
	\begin{equation}
		k_{i\mu}\sigma^{\mu}_{\alpha\dot{\beta}}=\slashed{k}_{i\alpha\dot{\beta}}=\sket{i}\bra{i}\qquad\epsilon^{\mu-1}_{i}=\frac{\sbra{r}\sigma^{\mu}\ket{i}}{\sbraket{ir}}\qquad\epsilon^{\mu+1}_{i}=\frac{\bra{r}\bar{\sigma}^{\mu}\sket{i}}{\braket{ri}},
	\end{equation}
	where $i$ denotes the particle label, $\mu$ denotes the Lorentz index. The $\pm 1$ on $\epsilon_{i}$ means the polarization of particle $i$ is $\pm 1$. $r$ denotes the reference spinor, which can be arbitrarily chosen to be any spinor not proportion to that of $i$. 
	\par Suppose we choose different reference spinor: $\sbra{r}=\sbra{2}$ or $\sbra{r}=\sbra{3}$ for $\epsilon^{\mu-1}_{i}$:
	\begin{equation}
		\epsilon^{\mu-1}_{i}=\frac{\sbra{2}\sigma^{\mu}\ket{i}}{\sbraket{i2}}\qquad\tilde{\epsilon}^{\mu-1}_{i}=\frac{\sbra{3}\sigma^{\mu}\ket{i}}{\sbraket{i3}},
	\end{equation}
	and compute the difference:
	\begin{equation}
		\epsilon^{\mu-1}_{i}-\tilde{\epsilon}^{\mu-1}_{i}=\frac{\sbra{2}\sigma^{\mu}\ket{i}}{\sbraket{i2}}-\frac{\sbra{3}\sigma^{\mu}\ket{i}}{\sbraket{i3}}=\frac{\sbra{2}\sigma^{\mu}\ket{i}\sbraket{i3}-\sbra{3}\sigma^{\mu}\ket{i}\sbraket{i2}}{\sbraket{i2}\sbraket{i3}},
	\end{equation}
	using simple Clifford algebra, we can conclude $\delta\epsilon^{\mu-1}_{i}= \epsilon^{\mu-1}_{i}-\tilde{\epsilon}^{\mu-1}_{i}\propto k^{\mu}_{i}$. This transformation on polarization is simply a linearized gauge transformation and does not have physical meaning. For convenience, we always choose the reference spinor to be the spinor of another particle and simplify our calculation. Take the four-point Yang-Mills amplitude $\mathcal{A}_{YM}(1^{+},2^{-},3^{-},4^{-})$ as an example. Since it breaks the MHV requirement, it is supposed to be $0$. We can choose the reference spinor of particle $2,3,4$ to be $\lambda_{1}$, and set $\lambda_{2}$ as the reference spinor of particle $1$, take $\epsilon_{2}\cdot\epsilon_{3}$ as an example:
	\begin{equation}
		\epsilon_{2}\cdot\epsilon_{3}=\frac{\sbra{1}\sigma^{\mu}\ket{2}}{\sbraket{21}}\frac{\sbra{1}\sigma^{\nu}\ket{3}}{\sbraket{31}}\eta_{\mu\nu}\propto\sbraket{11}=0,
	\end{equation}
this can be easily generalized to any dot products among polarization vectors\footnote{We will prove this relation for any two polarization vectors later on using Fierz identity \eqref{Fierz}.}. By momentum power counting of the numerators of $n$-point Yang-Mills amplitude, there is at most $n-2$ momentum. Together with $n$ polarization vectors of external legs contracting with metric, there must be at least one dot product between polarization vectors, which leads to the fact that amplitudes with all-minus helicity and single-plus helicity vanish.
\par Now that we can consider the interaction we are interested in, we will still start from a four-point example. 
\subsubsection{Pure gluon example}
The pure gluon example is also known as the Yang-Mills interaction, one of the cases in which the amplitude can be expressed using the closed form of spinor helicity. We can generate the current from the polarization. To calculate the amplitude more easily, we take the spinor helicity form. After applying the recursion rule of B-G currents and the spinor helicity form of the polarization vector, the B-G current that all on-shell gluons with the same helicity can be expressed as \cite{Berends:1987me}:
	\begin{equation}\label{bg1}
		\epsilon^{\mu}(i^{+},i+1^{+},\dots ,n^{+})=\frac{\bra{r}\bar{\sigma}^{\mu}\slashed{k}_{i,i+1,\dots,n}\ket{r}}{\sqrt{2}\braket{ri}\braket{i,i+1}\cdots\braket{n-1,n}\braket{nr}},
	\end{equation}
	where the $\bra{r}$ and $\ket{r}$ stand for the reference spinor, we have chosen the reference spinor for each on-shell leg to be the same. Using the recursion relation, we can get the B-G current where the first on-shell gluon has negative helicity. Here, we choose the reference spinor of particle $1$ to be $\lambda_{2}$, reference spinor for particle $2,3,\dots ,n$ as $\lambda_{1}$. We get \cite{Berends:1987me}:
	\begin{equation}\label{bg2}
		\epsilon^{\mu}(1^{-},2^{+},\dots ,n^{+})=\frac{\bra{1}\bar{\sigma}^{\mu}\slashed{k}_{2,3,\dots,n}\ket{1}}{\sqrt{2}\braket{12}\braket{23}\cdots\braket{n1}}\sum^{n}_{m=3}\frac{\bra{1}\slashed{k}_{m}\slashed{k}_{1,2,\dots,m}\ket{1}}{k^{2}_{1,2,\dots,m-1}k^{2}_{1,2,\dots,m}}.
	\end{equation}
	\par After contracting with the negative helicity gluon $n$ and some simple simplification, we have the well-known Parker-Taylor formula \cite{Parke:1986gb}:
	\begin{equation}\label{mhvyangmills}
		\mathcal{A}(1^{-},2^{+},3^{+},\cdots,n-1^{+},n^{-})=\frac{\braket{1n}^{4}}{\braket{12}\braket{23}\cdots\braket{n1}}.
	\end{equation}
	\subsubsection{Four-point QFT building block example}
	Consider the QFT building block of the four-point type-I super string $\mathcal{A}(1^{-},2^{+},3^{+}|\underline{4}^{-2})$, underline denotes the massive leg. The momentum of the particle is $k_{4}$, which can be decomposed into the summation of two null vectors, denoted by $a$ and $b$:
	\begin{equation}
		k^{\mu}_{4}=a^{\mu}+b^{\mu}\qquad a^{2}=b^{2}=0.
	\end{equation}
	We can always choose the reference spinor of particle $1$ to be $\lambda_{2}$, different reference spinor choices can always bring us different cancellations, but the final result of the amplitude is always the same.
	\par Let's set the reference spinor of particle $2,3$ to be $\lambda_{1}$. The cancellation table of the dot product between polarization vectors is:
	\begin{equation}\label{cancelation}
		\epsilon_{i}\dot\epsilon_{j}=0;\quad i.j=1,2,3
	\end{equation}
	The four-point string amplitude can be written as (eq.V.74) in \cite{Feng:2010yx}, as described in \eqref{QFTbuildingblock}, we focus on the QFT building block. In our reference spinor choice, we can easily find $\epsilon_{i}\cdot \epsilon_{j}=\epsilon_{2}\cdot k_{1}=\epsilon_{3}\cdot k_{1}=\epsilon_{1}\cdot k_{2}=0$, where $i,j=1,2,3$. We can simplify the four-point QFT building block to\footnote{We assume all constant factors are hidden in the disk integration building block, thus won't show up in our QFT amplitude.}:
	\begin{equation}
		\begin{aligned}
			\mathcal{A}&(1^{-},2^{+},3^{+}|\underline{4}^{-2})=\Phi_{\mu\nu}\times
			\\
			&\left\{
			(\epsilon_{3}\cdot k_{2})(\epsilon^{\mu}_{1}\epsilon^{\nu}_{2})-(\epsilon_{2}\cdot k_{3})(\epsilon^{\mu}_{1}\epsilon^{\nu}_{3})+(\epsilon_{1}\cdot k_{3})(\epsilon^{\mu}_{2}\epsilon^{\nu}_{3})
			+\frac{1}{s_{1,3}}(\epsilon_{1}\cdot k_{3})(\epsilon_{3}\cdot k_{2})(\epsilon^{\mu}_{2}k^{\nu}_{1})\right.
			\\
			&+\frac{1}{s_{1,3}}(\epsilon_{1}\cdot k_{3})(\epsilon_{3}\cdot k_{2})(\epsilon^{\mu}_{2}k^{\nu}_{3})+\frac{1}{s_{1,3}}(\epsilon_{1}\cdot k_{3})(\epsilon_{2}\cdot k_{3})(\epsilon^{\mu}_{3}k^{\nu})
			\\&\left.+\frac{1}{s_{2,3}}(\epsilon_{1}\cdot k_{3})(\epsilon_{3}\cdot k_{2})(\epsilon^{\mu}_{2}k^{\nu}_{1})-\frac{1}{s_{2,3}}(\epsilon_{1}\cdot k_{3})(\epsilon_{2}\cdot k_{3})(\epsilon^{\mu}_{3}k^{\nu}_{1})
			\right\},
		\end{aligned}
	\end{equation}
	we can now rewrite it into the spinor helicity form:
	\begin{equation}
		\mathcal{A}(1^{-},2^{+},3^{+}|\underline{4}^{-2})=\frac{\braket{1b}^{4}\sbraket{ab}^{2}}{m^{2}\braket{12}\braket{23}\braket{31}}.
	\end{equation}
	\par The central result of this work is a conjectural generalization to arbitrary $n$-point QFT building blocks:
	\begin{equation}\label{formula}
	    \mathcal{A}(1^{-},2^{+},\dots,n-2^{+},n-1^{+}|\underline{n}^{-2})=\frac{\sbraket{ab}^{2}}{2m^{2}}\frac{\braket{1b}^{4}}{\braket{12}\braket{23}\cdots\braket{n-1,1}},
	\end{equation}
    after applying the spin raising operator, we can easily get:
    \begin{equation}
	    \begin{aligned}
		\mathcal{A}(1^{-},2^{+},\dots,n-2^{+},n-1^{+}|\underline{n}^{-1})&=\frac{\sbraket{ab}^{2}}{m^{2}} \frac{\braket{1b}^{3}\braket{1a}}{\braket{12}\braket{23}\cdots\braket{n-1,1}},
		\\
		\mathcal{A}(1^{-},2^{+},\dots,n-2^{+},n-1^{+}|\underline{n}^{0})&=\frac{\sqrt{6}\sbraket{ab}^{2}}{2m^{2}} \frac{\braket{1b}^{2}\braket{1a}^{2}}{\braket{12}\braket{23}\cdots\braket{n-1,1}},
		\\
		\mathcal{A}(1^{-},2^{+},\dots,n-2^{+},n-1^{+}|\underline{n}^{+1})&=\frac{\sbraket{ab}^{2}}{m^{2}} \frac{\braket{1b}\braket{1a}^{3}}{\braket{12}\braket{23}\cdots\braket{n-1,1}},
		\\
		\mathcal{A}(1^{-},2^{+},\dots,n-2^{+},n-1^{+}|\underline{n}^{+2})&=\frac{\sbraket{ab}^{2}}{2m^{2}} \frac{\braket{1a}^{4}}{\braket{12}\braket{23}\cdots\braket{n-1,1}},
		\end{aligned}
	\end{equation}
	the aim of this paper is proving these formulas, for simplification, we will focus on the coupling with $n^{-2}$. Others can be generated by acting spin raising operator on the polarization.
	\subsubsection{Comparing helicity configurations}
	\par In last section, we gave the spinor helicity form of one specific helicity configuration \eqref{formula}, which is similar with the MHV helicity configuration in Yang-Mills theory \eqref{mhvyangmills}. The helicity configuration in \eqref{formula} will be refered to as MHV-like helicity configuration. There are also other helicity configurations, changing helicities in \eqref{formula} to all-plus would arrive at another QFT building block, which turns out to vanish:
	\begin{equation}
	    \mathcal{A}(1^{+},2^{+},\dots,n-2^{+},n-1^{+}|\underline{n}^{-2})=0.
	\end{equation}
	The vanishing of all-plus QFT building blocks to all multiplicities $n$ will be proven in appendix \ref{vanish} and is analogous to the vanishing of single-minus and all-plus helicity amplitudes in Yang-Mills theory. There are also helicity configurations similar with the NMHV, N$^{2}$MHV, $\dots$ in Yang-Mills theory, $\mathcal{A}(1^{-},2^{-},3^{+}\dots ,n-1^{+}|\underline{n})$ for example. The spinor helicity form of them are expected to be much more complicated in the same way as NMHV, N$^{2}$MHV, etc. helicity configurations in pure Yang-Mills theory give rise to more lengthy amplitude formulas than the MHV sector.
	\subsection{Basic idea of BCFW recursion}
	\par BCFW method \cite{Britto:2004ap,Britto:2005fq} aims to construct higher point QFT amplitudes using lower point QFT amplitudes. It is based on the factorization property of amplitudes\footnote{When the propagator becomes on-shell, the whole amplitude will decompose into products of two lower-multiplicity amplitudes.}, also known as the unitary requirement of amplitudes. The factorization property can be expressed as:

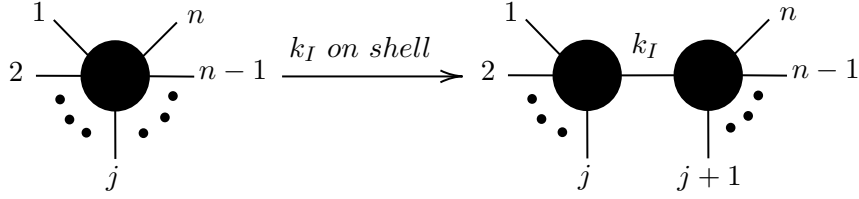
\tikzset{every picture/.style={line width=0.75pt}} 
\begin{center}
\begin{tikzpicture}[x=0.75pt,y=0.75pt,yscale=-1,xscale=1]

\draw    (369.03,169.47) -- (395.82,169.41) ;
\draw    (321.59,141.34) -- (338.72,159.13) ;
\draw    (312.69,169.41) -- (335.44,169.47) ;
\draw    (352.24,186.91) -- (352.32,211.26) ;
\draw    (443.01,141.34) -- (425.88,158.6) ;
\draw    (429.4,169.41) -- (451.91,169.59) ;
\draw    (412.61,186.85) -- (412.69,211.2) ;
\draw  [fill={rgb, 255:red, 0; green, 0; blue, 0 }  ,fill opacity=1 ][line width=0.75]  (322.85,181.31) .. controls (322.85,180.23) and (323.69,179.35) .. (324.74,179.35) .. controls (325.78,179.35) and (326.63,180.23) .. (326.63,181.31) .. controls (326.63,182.4) and (325.78,183.28) .. (324.74,183.28) .. controls (323.69,183.28) and (322.85,182.4) .. (322.85,181.31) -- cycle ;
\draw  [fill={rgb, 255:red, 0; green, 0; blue, 0 }  ,fill opacity=1 ][line width=0.75]  (327.55,191.43) .. controls (327.55,190.34) and (328.4,189.47) .. (329.44,189.47) .. controls (330.48,189.47) and (331.33,190.34) .. (331.33,191.43) .. controls (331.33,192.51) and (330.48,193.39) .. (329.44,193.39) .. controls (328.4,193.39) and (327.55,192.51) .. (327.55,191.43) -- cycle ;
\draw  [fill={rgb, 255:red, 0; green, 0; blue, 0 }  ,fill opacity=1 ][line width=0.75]  (337.96,198.05) .. controls (337.96,196.97) and (338.81,196.09) .. (339.85,196.09) .. controls (340.89,196.09) and (341.74,196.97) .. (341.74,198.05) .. controls (341.74,199.14) and (340.89,200.01) .. (339.85,200.01) .. controls (338.81,200.01) and (337.96,199.14) .. (337.96,198.05) -- cycle ;
\draw  [fill={rgb, 255:red, 0; green, 0; blue, 0 }  ,fill opacity=1 ][line width=0.75]  (435.7,179.57) .. controls (435.7,178.49) and (436.55,177.61) .. (437.59,177.61) .. controls (438.63,177.61) and (439.48,178.49) .. (439.48,179.57) .. controls (439.48,180.65) and (438.63,181.53) .. (437.59,181.53) .. controls (436.55,181.53) and (435.7,180.65) .. (435.7,179.57) -- cycle ;
\draw  [fill={rgb, 255:red, 0; green, 0; blue, 0 }  ,fill opacity=1 ][line width=0.75]  (430.66,189.86) .. controls (430.66,188.77) and (431.51,187.9) .. (432.55,187.9) .. controls (433.59,187.9) and (434.44,188.77) .. (434.44,189.86) .. controls (434.44,190.94) and (433.59,191.82) .. (432.55,191.82) .. controls (431.51,191.82) and (430.66,190.94) .. (430.66,189.86) -- cycle ;
\draw  [fill={rgb, 255:red, 0; green, 0; blue, 0 }  ,fill opacity=1 ][line width=0.75]  (421.93,195.79) .. controls (421.93,194.7) and (422.78,193.82) .. (423.82,193.82) .. controls (424.86,193.82) and (425.71,194.7) .. (425.71,195.79) .. controls (425.71,196.87) and (424.86,197.75) .. (423.82,197.75) .. controls (422.78,197.75) and (421.93,196.87) .. (421.93,195.79) -- cycle ;
\draw  [fill={rgb, 255:red, 0; green, 0; blue, 0 }  ,fill opacity=0.48 ] (99.94,169.72) .. controls (99.94,160.09) and (107.46,152.28) .. (116.74,152.28) .. controls (126.01,152.28) and (133.53,160.09) .. (133.53,169.72) .. controls (133.53,179.35) and (126.01,187.16) .. (116.74,187.16) .. controls (107.46,187.16) and (99.94,179.35) .. (99.94,169.72) -- cycle ;
\draw    (86.09,141.59) -- (103.22,159.38) ;
\draw    (77.19,169.66) -- (99.94,169.72) ;
\draw    (116.74,187.16) -- (116.82,211.51) ;
\draw  [fill={rgb, 255:red, 0; green, 0; blue, 0 }  ,fill opacity=1 ][line width=0.75]  (87.35,181.56) .. controls (87.35,180.48) and (88.19,179.6) .. (89.24,179.6) .. controls (90.28,179.6) and (91.13,180.48) .. (91.13,181.56) .. controls (91.13,182.65) and (90.28,183.53) .. (89.24,183.53) .. controls (88.19,183.53) and (87.35,182.65) .. (87.35,181.56) -- cycle ;
\draw  [fill={rgb, 255:red, 0; green, 0; blue, 0 }  ,fill opacity=1 ][line width=0.75]  (92.05,191.68) .. controls (92.05,190.59) and (92.9,189.72) .. (93.94,189.72) .. controls (94.98,189.72) and (95.83,190.59) .. (95.83,191.68) .. controls (95.83,192.76) and (94.98,193.64) .. (93.94,193.64) .. controls (92.9,193.64) and (92.05,192.76) .. (92.05,191.68) -- cycle ;
\draw  [fill={rgb, 255:red, 0; green, 0; blue, 0 }  ,fill opacity=1 ][line width=0.75]  (100.46,198.3) .. controls (100.46,197.22) and (101.31,196.34) .. (102.35,196.34) .. controls (103.39,196.34) and (104.24,197.22) .. (104.24,198.3) .. controls (104.24,199.39) and (103.39,200.26) .. (102.35,200.26) .. controls (101.31,200.26) and (100.46,199.39) .. (100.46,198.3) -- cycle ;
\draw    (147.51,142.84) -- (130.38,160.1) ;
\draw  [fill={rgb, 255:red, 0; green, 0; blue, 0 }  ,fill opacity=1 ][line width=0.75]  (139.66,190.61) .. controls (139.66,189.52) and (140.51,188.65) .. (141.55,188.65) .. controls (142.59,188.65) and (143.44,189.52) .. (143.44,190.61) .. controls (143.44,191.69) and (142.59,192.57) .. (141.55,192.57) .. controls (140.51,192.57) and (139.66,191.69) .. (139.66,190.61) -- cycle ;
\draw  [fill={rgb, 255:red, 0; green, 0; blue, 0 }  ,fill opacity=1 ][line width=0.75]  (128.93,198.54) .. controls (128.93,197.45) and (129.78,196.57) .. (130.82,196.57) .. controls (131.86,196.57) and (132.71,197.45) .. (132.71,198.54) .. controls (132.71,199.62) and (131.86,200.5) .. (130.82,200.5) .. controls (129.78,200.5) and (128.93,199.62) .. (128.93,198.54) -- cycle ;
\draw  [fill={rgb, 255:red, 0; green, 0; blue, 0 }  ,fill opacity=1 ][line width=0.75]  (143.66,179.86) .. controls (143.66,178.77) and (144.51,177.9) .. (145.55,177.9) .. controls (146.59,177.9) and (147.44,178.77) .. (147.44,179.86) .. controls (147.44,180.94) and (146.59,181.82) .. (145.55,181.82) .. controls (144.51,181.82) and (143.66,180.94) .. (143.66,179.86) -- cycle ;
\draw    (133.65,169.91) -- (156.16,170.09) ;
\draw  [fill={rgb, 255:red, 0; green, 0; blue, 0 }  ,fill opacity=0.48 ] (395.82,169.41) .. controls (395.82,159.78) and (403.33,151.98) .. (412.61,151.98) .. controls (421.88,151.98) and (429.4,159.78) .. (429.4,169.41) .. controls (429.4,179.04) and (421.88,186.85) .. (412.61,186.85) .. controls (403.33,186.85) and (395.82,179.04) .. (395.82,169.41) -- cycle ;
\draw  [fill={rgb, 255:red, 0; green, 0; blue, 0 }  ,fill opacity=0.48 ] (335.44,169.47) .. controls (335.44,159.84) and (342.96,152.03) .. (352,152.03) .. controls (361.51,152.03) and (369.03,159.84) .. (369.,169.47) .. controls (369.03,179.1) and (361.51,186.91) .. (352,186.91) .. controls (343,186.91) and (335.44,179.1) .. (335.44,169.47) -- cycle ;
\draw    (200,170) -- (288,170) ;
\draw [shift={(290,170)}, rotate = 180.16] [color={rgb, 255:red, 0; green, 0; blue, 0 }  ][line width=0.75]    (11,-3) .. controls (7,-1) and (3,0) .. (0,0) .. controls (3.31,0.3) and (7,1) .. (11,3)   ;

\draw (372.8,146.71) node [anchor=north west][inner sep=0.75pt]    {$k_{I}$};
\draw (309.12,131.12) node [anchor=north west][inner sep=0.75pt]    {$1$};
\draw (297.7,160.94) node [anchor=north west][inner sep=0.75pt]    {$2$};
\draw (346.24,213.08) node [anchor=north west][inner sep=0.75pt]    {$j$};
\draw (446.66,133.74) node [anchor=north west][inner sep=0.75pt]    {$n$};
\draw (452.78,159.99) node [anchor=north west][inner sep=0.75pt]    {$n-1$};
\draw (395.86,213.43) node [anchor=north west][inner sep=0.75pt]    {$j+1$};
\draw (73.62,131.37) node [anchor=north west][inner sep=0.75pt]    {$1$};
\draw (62.2,161.19) node [anchor=north west][inner sep=0.75pt]    {$2$};
\draw (110.74,213.33) node [anchor=north west][inner sep=0.75pt]    {$j$};
\draw (151.16,135.24) node [anchor=north west][inner sep=0.75pt]    {$n$};
\draw (157.03,160.49) node [anchor=north west][inner sep=0.75pt]    {$n-1$};
\draw (202.05,148.46) node [anchor=north west][inner sep=0.75pt]    {$k_{I} \ on\ shell$};

\end{tikzpicture}
\end{center}

	\begin{equation}\label{1.1}
		\mathcal{A}_{n}\xrightarrow{k_{I}\enspace on\enspace shell}\mathcal{A}_{L}\frac{1}{k_{I}}\mathcal{A}_{R}, 
	\end{equation}
	where $\mathcal{A}_{n}$ denotes an amplitude with $n$ external legs. While $\mathcal{A}_{L}$ and $\mathcal{A}_{R}$ denote sub-amplitudes on the left-hand side and right-hand side separately. Momentum $k_{I}$ is the internal momentum that flows from left to right or right to left, up to our choice. Suppose the external momentum of the left-hand side sub-amplitude is $k_{1},k_{2},\cdots k_{i}$, and define set $I$:
	\begin{equation}
		I=\{1,2,3,\cdots i\}.
	\end{equation}
	\par We set the direction of all $n$ external momentum outward and $k_{I}$ flow from left to right. After applying momentum conservation we have: $k_{I}=k_{1}+k_{2}+\cdots+k_{i}$. On the other hand, the external legs of $\mathcal{A}_{R}$ are denoted by set $J$:
	\begin{equation}
		J=\{i+1,i+2,\cdots n\},
	\end{equation}
	for the same reason, $k_{I}=-k_{i+1}-k_{i+2}-\cdots-k_{n}$.
	\par For most of the amplitudes we might need, $k_{I}$ is off shell\footnote{Which is easily seen using momentum conservation and the on-shell requirement for external momentum.}. After complex shifting external momentum:
	\begin{equation}
		\hat{k}^{\mu}_{i}=k^{\mu}_{i}+z r^{\mu}_{i}\qquad z\in\mathbb{C}\qquad i\in I,
	\end{equation}
	with some constraints on $r$, we can solve $z_{I}$ which makes $k_{I}$ on shell. The on-shell condition of external momentum and momentum conservation is preserved. $k_{I}$ is thus shifted as:
	\begin{equation}
		\hat{k}_{I}=k_{I}+z\sum_{i\in I} r_{i}.
	\end{equation}
	\par The complex shift $z r_{i}$ shouldn't influence the fundamental property of the amplitude and the shifted momentum. Thus, we want the external legs still satisfy their on-shell condition with mass unchanged. This will lead to some constraints on $r_{i}$. All $r_{i}$ satisfying the constraints can be the shift we use. To give a clean and convenient form of the expressions, we always choose the $r_{i}$ as simple as possible, and choose the suitable $z$, making $\hat{k}^{2}_{I}=0$ for some $I$ and factorize the amplitude we need into the multiplication of two sub-amplitudes and a pole on the $z$-plane. Each propagator corresponds to a pole on the complex plane. According to the Cauchy theorem, we have:
	\begin{equation}
		0=\sum_{z_{I}}Res_{z=z_{I}}\frac{\hat{\mathcal{A}}_{n}(z)}{z}-B_{n},
	\end{equation}
	where the index $I$ in $z_{I}$ denotes the pole corresponding to $k_{I}$ on-shell, and $B_{n}$ is the residual of the pole at the infinite point.
	\par Especially, $z_{I}=0$ stands for the amplitude without shifting external momentum, which is the original amplitude. We have:
	\begin{equation}
		\mathcal{A}_{n}=-\sum_{z_{I}\neq 0}Res_{z=z_{I}}\frac{\hat{\mathcal{A}}_{n}(z)}{z}+B_{n},
	\end{equation}
	\section{Result at five points}
In this section, we will take the five-point case as an example. Using the B-G recursion, we can analytically prove that the five-point QFT building blocks are:
	\begin{equation}\label{differentspin}
	    \begin{aligned}
		\mathcal{A}(1^{-1},2^{+1},3^{+1},4^{+1}|\underline{5}^{-2})&=\frac{\sbraket{ab}^{2}}{2m^{2}} \frac{\braket{1b}^{4}}{\braket{12}\braket{23}\braket{34}\braket{41}},
		\\
		\mathcal{A}(1^{-1},2^{+1},3^{+1},4^{+1}|\underline{5}^{-1})&=\frac{\sbraket{ab}^{2}}{m^{2}} \frac{\braket{1b}^{3}\braket{1a}}{\braket{12}\braket{23}\braket{34}\braket{41}},
		\\
		\mathcal{A}(1^{-1},2^{+1},3^{+1},4^{+1}|\underline{5}^{0})&=\frac{\sqrt{6}\sbraket{ab}^{2}}{2m^{2}} \frac{\braket{1b}^{2}\braket{1a}^{2}}{\braket{12}\braket{23}\braket{34}\braket{41}},
		\\
		\mathcal{A}(1^{-1},2^{+1},3^{+1},4^{+1}|\underline{5}^{+1})&=\frac{\sbraket{ab}^{2}}{m^{2}} \frac{\braket{1b}\braket{1a}^{3}}{\braket{12}\braket{23}\braket{34}\braket{41}},
		\\
		\mathcal{A}(1^{-1},2^{+1},3^{+1},4^{+1}|\underline{5}^{+2})&=\frac{\sbraket{ab}^{2}}{2m^{2}} \frac{\braket{1a}^{4}}{\braket{12}\braket{23}\braket{34}\braket{41}},
		\end{aligned}
	\end{equation}
	by using B-G currents, we construct the QFT building block for $\ket{-2,2}$ state coupling with four gluons and generate the coupling of other spins by applying the raising and lowering operator \eqref{rloperator} on polarization tensor $\Phi^{\mu\nu}$. 
	\par The underlined external leg stands for the massive leg. Here is particle $5$, we decompose the momentum $p^{\mu}_{5}$ into $p^{\mu}_{5}=a^{\mu}+b^{\mu}$, where $a$ and $b$ are both massless. After we work out the result for $\mathcal{A}(1^{-1},2^{+1},3^{+1},4^{+1}|\underline{5}^{-2})$, we would use the raising and lowering operator on the amplitude and get the result for different spin state of particle $5$. Thus, in this section, we only work out the coupling with $\ket{-2,2}$ state of particle $5$. For $\ket{-2,2}$ state, the polarization $\Phi^{\mu\nu}$ can be decomposed as:
	\begin{equation}
	    \Phi_{\mu\nu}^{(-2)}=\underline{\epsilon}^{(-1)}_{\mu}\underline{\epsilon}^{(-1)}_{\mu},
	\end{equation}
	the superscript $(-2)$ and $(-1)$ is just a index identifying the spin choice of the polarization, since we only analyze the coupling with spin $-2$ state of particle $5$, we will omit this index.
	\par According to the definition \eqref{amplitude}, we have:
	\begin{equation}\label{5pointamplitude}
		\mathcal{A}(1^{-1},2^{+1},3^{+1},4^{+1}|\underline{5}^{-2})= (\Phi_{1234})^{\mu\nu}(\Phi_{5})_{\mu\nu}=\sqrt{\ap} \sum_{P=QR}  F_{Q \, \rho}^{\mu}F_{R}^{\rho\nu}(\Phi_{5})_{\mu\nu}  +{\rm cyc}_P.
	\end{equation}
	where $P=1234$.
		\subsection{Some basic facts}\label{facts}
	\par Some equations can help us simplify our calculation. Before we move on, let's derive them first\footnote{These equations work for arbitrary $n$.}. 
	\par As before, we take the reference spinor of particle $1$ to be $\lambda_{2}$, and the reference spinor of all other massless particles\footnote{There is no need to define a reference spinor for massive particle.} $r_{2},r_{3}\dots, r_{j}=\lambda_{1}$. Thus, all polarization vector of massless particles have $\bra{1}$ or $\ket{1}$ in their numerator, by using the Fierz identities \eqref{Fierz}, we can easily find that the contraction between any two polarization vectors of a massless particle gives zero. This special choice of reference spinor can bring us more information. 
 \par First, we can express the Berends-Giele currents in terms of spinor helicities. Two examples are \eqref{bg1} and \eqref{bg2}. With our reference spinor choice, we rewrite the closed form of the B-G current for massless states in the QFT building block of type-I superstring as follows:
	\begin{equation}\label{shform}
		\begin{aligned}
			\left.\epsilon^{\mu}(i^{+},i+1^{+},\dots ,n^{+})\right|_{r_{i,i+1,\dots,n}^{\mu}=k_{1}^{\mu}}&=\frac{\bra{1}\bar{\sigma}^{\mu}\slashed{k}_{i,i+1,\dots,n}\ket{1}}{\sqrt{2}\braket{1i}\braket{i,i+1}\cdots\braket{n-1,n}\braket{n1}}
			\\
			\left.\epsilon^{\mu}(1^{-},2^{+},\dots ,n^{+})\right|_{r_{2,3,\dots,n}^{\mu}=k_{1}^{\mu},r_{1}=k_{2}^{\mu}}&=\frac{\bra{1}\bar{\sigma}^{\mu}\slashed{k}_{2,3,\dots,n}\ket{1}}{\sqrt{2}\braket{12}\braket{23}\cdots\braket{n1}}\sum^{n}_{m=3}\frac{\bra{1}\slashed{k}_{m}\slashed{k}_{1,2,\dots,m}\ket{1}}{k^{2}_{1,2,\dots,m-1}k^{2}_{1,2,\dots,m}}
		\end{aligned}
	\end{equation}
	under this reference spinor choice, using B-G recursion, we can easily show that the spinor helicity closed form of B-G current always have a similar numerator structure $\bra{1}\bar{\sigma}^{\mu}\slashed{k}_{i,j}\ket{1}$. Therefore, if we contract any two B-G currents, using Fierz identity \eqref{Fierz}, we immediately get a result proportional to $\braket{11}=0$.
	\par Second, let's contract any B-G current with momentum $k^{\mu}_{1}$, we can always get a result proportional to $\bra{1}\slashed{k}_{1}$. We can expand $\slashed{k}_{1}=-\ket{1}\sbra{1}$. The final result is proportional to $\braket{11}=0$. For the same reason, $\epsilon_{1}\cdot k_{2}=0$.
	\par Third, the B-G current we use is the same as the one for Yang-Mills amplitude, which is: $\epsilon^{\mu}_{12}|_{r_{2}^{\mu}=k_{1}^{\mu},r_{1}=k_{2}^{\mu}}=\epsilon^{\mu}_{21}|_{r_{2}^{\mu}=k_{1}^{\mu},r_{1}=k_{2}^{\mu}}=0$:
	\begin{equation}
	    \begin{aligned}
		\left.\epsilon^{\mu}_{12}\right|_{r_{2}^{\mu}=k_{1}^{\mu},r_{1}=k_{2}^{\mu}}&=\left.\frac{1}{2s_{12}}[(k_{2}\cdot \epsilon_{1})\epsilon^{\mu}_{2}-(k_{1}\cdot \epsilon_{2})\epsilon^{\mu}_{1}]\right|_{r_{2}^{\mu}=k_{1}^{\mu},r_{1}=k_{2}^{\mu}}
		\\
		&=\left.\frac{1}{2s_{12}}[(k_{2}\cdot \epsilon_{1})\epsilon^{\mu}_{2}-(k_{1}\cdot\epsilon_{2})\epsilon^{\mu}_{1}]\right|_{r_{2}^{\mu}=k_{1}^{\mu},r_{1}=k_{2}^{\mu}}=0
	    \end{aligned}
	\end{equation}
	\subsection{Contributions of each configuration}
	\par In \eqref{5pointamplitude}, $P=1,2,3,4$ are divided into $Q$ and $R$. Thus, we can write down all the configurations of $Q$ and $R$ before we calculate each of them\footnote{Notice $(\Phi_{5})_{\mu\rho}$ is symmetric in $m$ and $p$ indices. Thus we can exchange $Q$ and $R$ without changing the final result. This decreases the number of configurations by a factor of two}:
	\begin{equation}\label{configurations}
		\begin{tabular}{|c|c|c|c|c|c|c|}
			\hline configurations&1&2&3&4&5&6\\
			\hline$Q$&1&1,2&1,2,3&2&2,3&3\\
			\hline$R$&2,3,4&3,4&4&3,4,1&4,1&4,1,2\\
			\hline
		\end{tabular}
	\end{equation}

	\subsubsection{The only non-zero configuration}\label{1;2,n}
	\par Among all $6$ configurations in \eqref{configurations}, there is only one configuration that nontrivially contributes to the final result of the 5-pt QFT building block, which is the configuration 1. Consider the coupling with $\ket{-2,2}$ state for computation simplicity:
	\begin{equation}
		\begin{aligned}
			&F_{1 \, \rho}^{\mu}F_{234}^{\rho\nu}(\Phi_{5})_{\mu\nu}
			\\
			=& (k_1^{\mu} \ep_{1\rho} - k_{1\rho} \ep_1^{\mu})[k_{234}^{\rho} \ep_{234}^{\nu} - k_{234}^{\nu} \ep_{234}^{\rho} - (\ep_2^{\rho} \ep_{34}^{\nu} - \ep_2^{\nu} \ep_{34}^{\rho}+\ep_{23}^{\rho} \ep_4^{\nu} - \ep_{23}^{\nu} \ep_4^{\rho})](\Phi_{5})_{\mu\nu}
			\\
			=& (k_1^{\mu} \ep_{1\rho} - k_{1\rho} \ep_1^{\mu})[k_{234}^{\rho} \ep_{234}^{\nu} - k_{234}^{\nu} \ep_{234}^{\rho} ](\Phi_{5})_{\mu\nu}
		\end{aligned}
	\end{equation}
	where we applied the basic rules, we concluded in section \ref{facts} in the second step. After converting into the spinor helicity form and applying \eqref{bg1}, we get:
	\begin{equation}
		\begin{aligned}
			&F_{1 \, \rho}^{\mu}F_{234}^{\rho\nu}(\Phi_{5})_{\mu\nu}
			\\
			=& \frac{\bra{b}\slashed{k}_{1}\sket{a}}{\sqrt{2}m}\frac{\sbra{2}\slashed{k}_{234}\ket{1}}{\sqrt{2}\sbraket{12}}\frac{\braket{1b}\sbra{a}\slashed{k}_{234}\ket{1}}{m\braket{12}\braket{23}\braket{34}\braket{41}}+2\frac{\sbraket{2a}\braket{b1}}{2\sbraket{12}m}(k_{1}\cdot k_{234})\frac{\braket{1b}\sbra{a}\slashed{k}_{234}\ket{1}}{m\braket{12}\braket{23}\braket{34}\braket{41}}
			\\
			=&\frac{\braket{1b}^{4}\sbraket{ab}^{2}}{2m^{2}\braket{12}\braket{23}\braket{34}\braket{41}}
		\end{aligned}
	\end{equation}
	where this is exactly the $5$-pt amplitude, we claimed.
	\subsubsection{Other configurations vanish individually}
	\par We can analytically prove that except for the configuration mentioned in section \ref{1;2,n}, all other $5$ configurations vanish individually. We will show these one by one:
	\begin{itemize}
		\item Configuration $2$
		\par Configuration $2$, $F_{12 \, \rho}^{\mu}F_{34}^{\rho\nu}(\Phi_{5})_{\mu\nu}$ equals to:
		\begin{equation}
			\begin{aligned}
				&F_{12 \, \rho}^{\mu}F_{34}^{\rho\nu}(\Phi_{5})_{\mu\nu}
				\\
				=&[k_{12}^{\mu} \ep_{12\rho} - k_{12\rho} \ep_{12}^{\mu} - (\ep_{1}^{\mu} \ep_{2\rho} - \ep_{1\rho} \ep_{2}^{\mu})][k_{34}^{\rho} \ep_{34}^{\nu} - k_{34}^{\nu} \ep_{34}^{\rho} - (\ep_{3}^{\rho} \ep_{4}^{\nu} - \ep_{3}^{\nu} \ep_{4}^{\rho})](\Phi_{5})_{\mu\nu}
				\\
				=& - (\ep_{1}^{\mu} \ep_{2\rho} - \ep_{1\rho} \ep_{2}^{\mu})k_{34}^{\rho} \ep_{34}^{\nu}(\Phi_{5})_{\mu\nu}=0
			\end{aligned}
		\end{equation}
		\item Configuration $3$
		\par Configuration $3$, $F_{123 \, \rho}^{\mu}F_{4}^{\rho\nu}(\Phi_{5})_{\mu\nu}$ equals to:
		\begin{equation}
			\begin{aligned}
				&F_{123 \, \rho}^{\mu}F_{4}^{\rho\nu}(\Phi_{5})_{\mu\nu}
				\\
				=&[k_{123}^{\mu} \ep_{123\rho} - k_{123\rho} \ep_{123}^{\mu} - (\ep_{1}^{\mu} \ep_{23\rho} - \ep_{1\rho} \ep_{23}^{\mu}+\ep_{12}^{\mu} \ep_{3\rho} - \ep_{12\rho} \ep_{3}^{\mu})][k_{4}^{\rho} \ep_{4}^{\nu} - k_{4}^{\nu} \ep_{4}^{\rho}](\Phi_{5})_{\mu\nu}
				\\
				=&-\frac{4\braket{1b}^{2}\sbraket{24}\left(k_{14}\cdot k_{23}\bra{1}\slashed{k}_{23}\sket{a}\sbraket{4a}+\bra{1}\slashed{k}_{23}\sket{4}\bra{b}\slashed{k}_{14}\sket{a}\sbraket{ba}\right)}{m^{2}\braket{12}\braket{13}\braket{14}\braket{23}\sbraket{14}\sbraket{21}}
				\\
				&+\frac{4\braket{b1}^{2}\sbraket{4a}^{2}\sbraket{24}\sbraket{23}}{m^{2}\braket{12}\braket{13}\sbraket{14}\sbraket{21}}-\frac{4\braket{b1}^{2}\sbraket{24}\bra{1}\slashed{k}_{23}\sket{a}^{2}}{m^{2}\braket{12}\braket{13}\braket{14}\braket{23}\sbraket{21}}=0
			\end{aligned}
		\end{equation}
		\item Configuration $4$
		\par Configuration $4$, $F_{2 \, \rho}^{\mu}F_{341}^{\rho\nu}(\Phi_{5})_{\mu\nu}$ equals to:
		\begin{equation}
			\begin{aligned}
				&F_{2 \, \rho}^{\mu}F_{341}^{\rho\nu}(\Phi_{5})_{\mu\nu}
				\\
				=&[k_{2}^{\mu} \ep_{2\rho} - k_{2\rho} \ep_{2}^{\mu}][k_{341}^{\rho} \ep_{341}^{\nu} - k_{341}^{\nu} \ep_{341}^{\rho} - (\ep_{34}^{\rho} \ep_{1}^{\nu} - \ep_{34}^{\nu} \ep_{1}^{\rho}+\ep_{3}^{\rho} \ep_{41}^{\nu} - \ep_{3}^{\nu} \ep_{41}^{\rho})](\Phi_{5})_{\mu\nu}
				\\
				=&-\frac{4\braket{1b}^{2}\sbraket{2a}^{2}\left(\braket{13}\sbraket{14}\sbraket{32}+2k_{4}\cdot k_{13}\sbraket{42}\right)}{m^{2}\braket{12}\braket{23}\braket{34}\braket{41}\sbraket{14}\sbraket{21}}
				\\
				&+\frac{4\braket{1b}^{2}\sbraket{2a}^{2}\left(\sbraket{41}\bra{1}\slashed{k}_{34}\sket{2}+2k_{3}\cdot k_{4}\sbraket{42}\right)}{m^{2}\braket{12}\braket{23}\braket{34}\braket{41}\sbraket{14}\sbraket{21}}=0
			\end{aligned}
		\end{equation}
		\item Configuration $5$
		\par Configuration $5$, $F_{23 \, \rho}^{\mu}F_{41}^{\rho\nu}(\Phi_{5})_{\mu\nu}$ equals to:
		\begin{equation}
			\begin{aligned}
				&F_{23 \, \rho}^{\mu}F_{41}^{\rho\nu}(\Phi_{5})_{\mu\nu}
				\\
				=&[k_{23}^{\mu} \ep_{23\rho} - k_{23\rho} \ep_{23}^{\mu} - (\ep_{2}^{\mu} \ep_{3\rho} - \ep_{2\rho} \ep_{3}^{\mu})][k_{41}^{\rho} \ep_{41}^{\nu} - k_{41}^{\nu} \ep_{41}^{\rho} - (\ep_{4}^{\rho} \ep_{1}^{\nu} - \ep_{4}^{\nu} \ep_{1}^{\rho})](\Phi_{5})_{\mu\nu}
				\\
				=& -\frac{4\braket{1b}^{2}\sbraket{24}\left(\bra{1}\slashed{k}_{23}\sket{4}\bra{b}\slashed{k}_{23}\sket{a}\sbraket{ab}+k_{14}\cdot k_{23}\bra{1}\slashed{k}_{23}\sket{a}\sbraket{4a}\right)}{m^{2}\braket{12}\braket{13}\braket{14}\braket{23}\sbraket{14}\sbraket{21}}
				\\
				&+\frac{4\braket{b1}^{2}\sbraket{4a}^{2}\sbraket{24}\sbraket{23}}{m^{2}\braket{12}\braket{13}\sbraket{14}\sbraket{21}}-\frac{4\braket{b1}^{2}\sbraket{24}\bra{1}\slashed{k}_{23}\sket{a}^{2}}{m^{2}\braket{12}\braket{13}\braket{14}\braket{23}\sbraket{21}}=0
			\end{aligned}
		\end{equation}
		\item Configuration $6$
		\par Configuration $6$, $F_{3 \, \rho}^{\mu}F_{412}^{\rho\nu}(\Phi_{5})_{\mu\nu}$ equals to:
		\begin{equation}
			\begin{aligned}
				&F_{3 \, \rho}^{\mu}F_{412}^{\rho\nu}(\Phi_{5})_{\mu\nu}
				\\
				=& (k_3^{\mu} \ep_{3\rho} - k_{3\rho} \ep_3^{\mu})[k_{412}^{\rho} \ep_{412}^{\nu} - k_{412}^{\nu} \ep_{412}^{\rho} - (\ep_4^{\rho} \ep_{12}^{\nu} - \ep_4^{\nu} \ep_{12}^{\rho}+\ep_{41}^{\rho} \ep_2^{\nu} - \ep_{41}^{\nu} \ep_2^{\rho})](\Phi_{5})_{\mu\nu}
				\\
				=&-\frac{4  \braket{1b}^2  \sbraket{24}^2 \sbraket{3a} (\bra{1}\slashed{k}_{24}\sket{a} k_{3}\cdot k_{5} + 
					\bra{1}\slashed{k}_{24}\sket{3} \bra{3}\slashed{b}\sket{a})}{m^2 k^{2}_{12}k^{2}_{14} \braket{13}  k^{2}_{124}}
				\\
				&+\frac{4 \braket{1b}^2 \sbraket{24}^2 \sbraket{3a}^2}{m^2 k^{2}_{12}k^{2}_{14}}=0 
			\end{aligned}
		\end{equation}
	\end{itemize}
	\section{The n-pt amplitude}
	\par One can easily apply \eqref{2.31}, \eqref{2.32} and the definition of $\Phi_{P}$ \eqref{currentsfieldstrengthandgauge} for multiple particles to the $n$-pt amplitude \eqref{amplitude} and count the number of terms in the B-G recursions. More specifically, we shall count the number of configurations $F^{\mu}_{Q\rho}F^{\rho\nu}_{R}$ compatible with the cyclic ordering of $\mathcal{A}(1,2,\dots,n-1|\underline{n})$. Similar to \eqref{configurations}, the number of configurations for $n$-pt amplitude is $\frac{n^{2}-3n+2}{2}$, which increases at order of $O(n^{2})$. From the computation of 5-pt, one can also tell that the complexity of each configuration increases when considering higher point amplitudes. Thus we can't analytically work out the closed form one by one. However, there is some special configuration that we can easily work out for arbitrary $n$ point, one of them equals to the $n$ point QFT building block we claimed:
	\begin{equation}\label{expected}
	    \mathcal{A}(\hat{1}^{-},2^{+},\dots,n-2^{+},n-1^{+}|\underline{\hat{n}}^{-2})=\frac{\sbraket{ab}^{2}}{2m^{2}}\frac{\braket{1b}^{4}}{\braket{12}\braket{23}\cdots\braket{n-1,1}}
	\end{equation}
	
	\subsection{The non zero configuration}\label{npt}
	\par Similar to the $5$-pt example, one configuration contributes to the closed form we expected for $n$-pt amplitude:
	\begin{equation}
		\begin{aligned}
			&F_{1 \, \rho}^{\mu}F_{23\dots n-1}^{\rho\nu}(\Phi_{n})_{\mu\nu}
			\\
			=& (k_1^{\mu} \ep_{1p} - k_{1p} \ep_1^{\mu})[k_{23\dots n-1}^{\rho} \ep_{23\dots n-1}^{\nu} - k_{23\dots n-1}^{\nu} \ep_{23\dots n-1}^{\rho} - (\ep_2^{\rho} \ep_{3\dots n}^{\nu} - \ep_2^{\nu} \ep_{3\dots n}^{\rho}+\cdots)](\Phi_{n})_{\mu\nu}
			\\
			=& (k_1^{\mu} \ep_{1p} - k_{1p} \ep_1^{\mu})[k_{23\dots n-1}^{\rho} \ep_{23\dots n-1}^{\nu} - k_{23\dots n-1}^{\nu} \ep_{23\dots n-1}^{\rho}](\Phi_{n})_{\mu\nu},
		\end{aligned}
	\end{equation}
	where in the last step, we used the important facts in section \ref{facts} that any two B-G currents contract to zero, and $k_{1}$ contracting with any B-G currents gives zero. After applying the spinor helicity form, the QFT building block of $\ket{-2,2}$ state coupling with $n-1$ gluons equals to:
	\begin{equation}\label{nonzeroconf}
		\begin{aligned}
		    F_{1 \, \rho}^{\mu}F_{23\dots n-1}^{\rho\nu}(\Phi_{n})_{\mu\nu}=& \frac{\bra{b}\slashed{k}_{1}\sket{a}}{\sqrt{2}m}\frac{\sbra{2}\slashed{k}_{23\dots n-1}\ket{1}}{\sqrt{2}\sbraket{12}}\frac{\braket{1b}\sbra{a}\slashed{k}_{23\dots n-1}\ket{1}}{m\braket{12}\braket{23}\cdots\braket{n-2,n-1}\braket{n-1,1}}
			\\
			&+2\frac{\sbraket{2a}\braket{b1}}{2\sbraket{12}m}(k_{1}\cdot k_{23\dots n-1})\frac{\braket{1b}\sbra{a}\slashed{k}_{23\dots n-1}\ket{1}}{m\braket{12}\braket{23}\cdots\braket{n-2,n-1}\braket{n-1,1}}
			\\
			=&\frac{\sbraket{ab}^{2}}{2m^{2}}\frac{\braket{1b}^{4}}{\braket{12}\braket{23}\cdots\braket{n-2,n-1}\braket{n-1,1}}
		\end{aligned}
	\end{equation}
	where we used Schouten identity, momentum conservation, and Fierz identity. It is exactly what we claimed in \eqref{formula}
\subsection{Other configurations}
For other configurations, we believe they are individually equal to zero as it goes for $5$-pt amplitude. We can prove one of them equals zero analytically, but for others, what we can do so far is numerically check up to $6$-pt. This numerical check provided the confirmation of \eqref{formula} at $n=6$. Similarly, one can give a numerical check for $n=7$ or higher point QFT building block.
\par We can prove that one configuration goes to zero individually for $n$-pt amplitude. This is a generalization of configuration $2$ in \eqref{configurations}. More specifically, we divide $1,2,3,\dots,n$ into $1,2$ and $3,\dots,n$. Still, we focus on the QFT building block of spin $-2$ massive particle coupling with gluons. 
\begin{equation}
\begin{aligned}
    &F_{12 \, \rho}^{\mu}F_{3\dots n-1}^{\rho\nu}(\Phi_{n})_{\mu\nu}
    \\
    =&[k_{12}^{\mu} \ep_{12\rho} - k_{12\rho} \ep_{12}^{\mu} - (\ep_{1}^{\mu} \ep_{2\rho} - \ep_{1\rho} \ep_{2}^{\mu})]
    \\
    &\times[k_{3\dots,n-1}^{\rho} \ep_{3\dots,n-1}^{\nu} - k_{3\dots,n-1}^{\nu} \ep_{3\dots,n-1}^{\rho} - (\ep_{3}^{\rho} \ep_{4\dots,n}^{\nu} - \ep_{3}^{\nu} \ep_{4\dots,n}^{\rho}+\cdots)](\Phi_{n})_{\mu\nu}
    \\
    =& - (\ep_{1}^{\mu} \ep_{2\rho} - \ep_{1\rho} \ep_{2}^{\mu})k_{3\dots,n-1}^{\rho} \ep_{3\dots,n-1}^{\nu}(\Phi_{n})_{\mu\nu}
    \\
    \propto&\left(\frac{\braket{1b} \sbraket{a2}\bra{1}\slashed{k}_{3\dots,n-1}\sket{2}}{k^{2}_{12}}-\frac{\bra{1}\slashed{k}_{3\dots,n-1}\sket{2}\braket{1b}\sbraket{a2}}{k^{2}_{12}}\right)
    \\
    =&0,
\end{aligned}
\end{equation}
and as for other configurations, we can numerically check they vanish separately for 6-point amplitude. Thus we believe \eqref{nonzeroconf} is the only non-zero configuration that contributes to the expected result for arbitrary n-point and all other configurations sum to zero. We will later on prove this using the BCFW recursion. 
\section{Application of BCFW recursion on QFT building block}
Considering the helicity configuration, the naive BCFW shifted leg should be $1$ and $n$, the external leg $n$ is massive, thus the BCFW shift for the massive particle is needed. We will show in the appendix that this kind of shift can indeed provide us with the expected result, but exists a tentative boundary term. In this section we will show if we shift leg $2$ and $3$ instead, we can still find the expected result without any BCFW boundary. 
\par The momentum is shifted as:
\begin{equation}
    \hat{k}^{\mu}_{2}=k^{\mu}_{2}+z r^{\mu} \qquad\hat{k}^{\mu}_{3}=k^{\mu}_{3}-z r^{\mu}
\end{equation}
where the $\slashed{r}$ needs to satisfy was choosen to be:
\begin{equation}
    \slashed{r}=\sket{3}\bra{2}+\ket{2}\sbra{3}
\end{equation}
\par Shifting the momentum leads to the following shifted spinor helicities:
\begin{equation}
\begin{aligned}
    \sket{\hat{2}}&=\sket{2}+z\sket{3}
    \\
    \ket{\hat{3}}&=\ket{3}-z\ket{2}
\end{aligned}
\end{equation}
where other spinor helicity is unshifted. We have the following relations:
\begin{equation}
    \braket{\hat{2}\hat{3}}= \braket{23}\qquad\sbraket{\hat{2}\hat{3}}=\braket{23}.
\end{equation}
We can now rewrite the four-point amplitude into spinor helicity form:
\begin{equation}
    \mathcal{A}(1^{-},2^{+},3^{+}|\underline{4}^{-2})=\frac{\sbraket{ab}^{2}}{2m^{2}}\frac{\braket{1b}^{4}}{\braket{12}\braket{23}\braket{31}},
\end{equation}
to prove the expected form, we need to assume the $n-1$ point amplitudes satisfy:
\begin{equation}
    \mathcal{A}(1^{-},2^{+},k_{I}^{+},5^{+},\dots,n-2^{+},n-1^{+}|\underline{n}^{-2})=\frac{\sbraket{ab}^{2}}{2m^{2}}\frac{\braket{1b}^{4}}{\braket{12}\braket{2k_{I}}\braket{k_{I}5}\cdots\braket{n-1,1}},
\end{equation}
and show that the existence of the expected form for any $n-1$ point always leads to the existence of $n$ point, since we already proved the five-point case using B-G recursion, if such $n-1\to n$ always exists, such form would be true for arbitrary $n$ point amplitude. The shifted $n$ point amplitude equals the sum of all residues of poles on the $z$ plane, which corresponds to the sum of different configurations of subamplitudes.
\par However, one can easily see that there exists only one pole which gives a nonvanishing contribution when we shift $2$ and $3$. Which is:

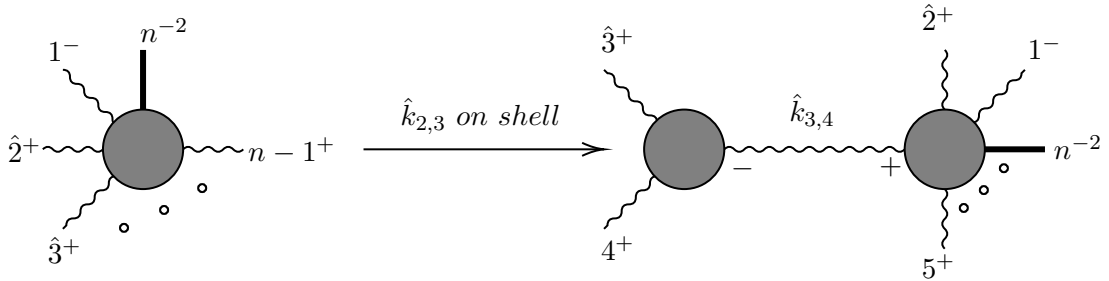
\tikzset{every picture/.style={line width=0.75pt}} 

\begin{tikzpicture}[x=0.75pt,y=0.75pt,yscale=-1,xscale=1]

\draw    (390,150) .. controls (391.67,148.33) and (393.33,148.33) .. (395,150) .. controls (396.67,151.67) and (398.33,151.67) .. (400,150) .. controls (401.67,148.33) and (403.33,148.33) .. (405,150) .. controls (406.67,151.67) and (408.33,151.67) .. (410,150) .. controls (411.67,148.33) and (413.33,148.33) .. (415,150) .. controls (416.67,151.67) and (418.33,151.67) .. (420,150) .. controls (421.67,148.33) and (423.33,148.33) .. (425,150) .. controls (426.67,151.67) and (428.33,151.67) .. (430,150) .. controls (431.67,148.33) and (433.33,148.33) .. (435,150) .. controls (436.67,151.67) and (438.33,151.67) .. (440,150) .. controls (441.67,148.33) and (443.33,148.33) .. (445,150) .. controls (446.67,151.67) and (448.33,151.67) .. (450,150) .. controls (451.67,148.33) and (453.33,148.33) .. (455,150) .. controls (456.67,151.67) and (458.33,151.67) .. (460,150) .. controls (461.67,148.33) and (463.33,148.33) .. (465,150) .. controls (466.67,151.67) and (468.33,151.67) .. (470,150) .. controls (471.67,148.33) and (473.33,148.33) .. (475,150) .. controls (476.67,151.67) and (478.33,151.67) .. (480,150) -- (480,150) ;
\draw    (500,100) .. controls (501.67,101.67) and (501.67,103.33) .. (500,105) .. controls (498.33,106.67) and (498.33,108.33) .. (500,110) .. controls (501.67,111.67) and (501.67,113.33) .. (500,115) .. controls (498.33,116.67) and (498.33,118.33) .. (500,120) .. controls (501.67,121.67) and (501.67,123.33) .. (500,125) .. controls (498.33,126.67) and (498.33,128.33) .. (500,130) -- (500,130) ;
\draw    (500,170) .. controls (501.67,171.67) and (501.67,173.33) .. (500,175) .. controls (498.33,176.67) and (498.33,178.33) .. (500,180) .. controls (501.67,181.67) and (501.67,183.33) .. (500,185) .. controls (498.33,186.67) and (498.33,188.33) .. (500,190) .. controls (501.67,191.67) and (501.67,193.33) .. (500,195) .. controls (498.33,196.67) and (498.33,198.33) .. (500,200) -- (500,200) ;
\draw    (120,150) .. controls (121.67,148.33) and (123.33,148.33) .. (125,150) .. controls (126.67,151.67) and (128.33,151.67) .. (130,150) .. controls (131.67,148.33) and (133.33,148.33) .. (135,150) .. controls (136.67,151.67) and (138.33,151.67) .. (140,150) .. controls (141.67,148.33) and (143.33,148.33) .. (145,150) .. controls (146.67,151.67) and (148.33,151.67) .. (150,150) -- (150,150) ;
\draw    (80,150) .. controls (78.33,151.67) and (76.67,151.67) .. (75,150) .. controls (73.33,148.33) and (71.67,148.33) .. (70,150) .. controls (68.33,151.67) and (66.67,151.67) .. (65,150) .. controls (63.33,148.33) and (61.67,148.33) .. (60,150) .. controls (58.33,151.67) and (56.67,151.67) .. (55,150) .. controls (53.33,148.33) and (51.67,148.33) .. (50,150) -- (50,150) ;
\draw    (60,110) .. controls (62.36,110) and (63.54,111.18) .. (63.54,113.54) .. controls (63.54,115.89) and (64.72,117.07) .. (67.07,117.07) .. controls (69.43,117.07) and (70.61,118.25) .. (70.61,120.61) .. controls (70.61,122.96) and (71.79,124.14) .. (74.14,124.14) .. controls (76.5,124.14) and (77.68,125.32) .. (77.68,127.68) .. controls (77.68,130.03) and (78.86,131.21) .. (81.21,131.21) .. controls (83.57,131.21) and (84.75,132.39) .. (84.75,134.75) .. controls (84.75,137.1) and (85.93,138.28) .. (88.28,138.28) -- (90,140) -- (90,140) ;
\draw    (90,160) .. controls (90,162.36) and (88.82,163.54) .. (86.46,163.54) .. controls (84.11,163.54) and (82.93,164.72) .. (82.93,167.07) .. controls (82.93,169.43) and (81.75,170.61) .. (79.39,170.61) .. controls (77.04,170.61) and (75.86,171.79) .. (75.86,174.14) .. controls (75.86,176.5) and (74.68,177.68) .. (72.32,177.68) .. controls (69.97,177.68) and (68.79,178.86) .. (68.79,181.21) .. controls (68.79,183.57) and (67.61,184.75) .. (65.25,184.75) .. controls (62.9,184.75) and (61.72,185.93) .. (61.72,188.28) -- (60,190) -- (60,190) ;
\draw  [fill={rgb, 255:red, 128; green, 128; blue, 128 }  ,fill opacity=1 ] (80,150) .. controls (80,138.95) and (88.95,130) .. (100,130) .. controls (111.05,130) and (120,138.95) .. (120,150) .. controls (120,161.05) and (111.05,170) .. (100,170) .. controls (88.95,170) and (80,161.05) .. (80,150) -- cycle ;
\draw [line width=2.25]    (100,130) -- (100,100) ;
\draw  [line width=3]  (90,189.5) .. controls (90,189.22) and (90.22,189) .. (90.5,189) .. controls (90.78,189) and (91,189.22) .. (91,189.5) .. controls (91,189.78) and (90.78,190) .. (90.5,190) .. controls (90.22,190) and (90,189.78) .. (90,189.5) -- cycle ;
\draw  [line width=3]  (110,180.5) .. controls (110,180.22) and (110.22,180) .. (110.5,180) .. controls (110.78,180) and (111,180.22) .. (111,180.5) .. controls (111,180.78) and (110.78,181) .. (110.5,181) .. controls (110.22,181) and (110,180.78) .. (110,180.5) -- cycle ;
\draw  [line width=3]  (129,169.5) .. controls (129,169.22) and (129.22,169) .. (129.5,169) .. controls (129.78,169) and (130,169.22) .. (130,169.5) .. controls (130,169.78) and (129.78,170) .. (129.5,170) .. controls (129.22,170) and (129,169.78) .. (129,169.5) -- cycle ;
\draw    (210,150) -- (328,150) ;
\draw [shift={(330,150)}, rotate = 180] [color={rgb, 255:red, 0; green, 0; blue, 0 }  ][line width=0.75]    (10.93,-3.29) .. controls (6.95,-1.4) and (3.31,-0.3) .. (0,0) .. controls (3.31,0.3) and (6.95,1.4) .. (10.93,3.29)   ;
\draw    (360,160) .. controls (360,162.36) and (358.82,163.54) .. (356.46,163.54) .. controls (354.11,163.54) and (352.93,164.72) .. (352.93,167.07) .. controls (352.93,169.43) and (351.75,170.61) .. (349.39,170.61) .. controls (347.04,170.61) and (345.86,171.79) .. (345.86,174.14) .. controls (345.86,176.5) and (344.68,177.68) .. (342.32,177.68) .. controls (339.97,177.68) and (338.79,178.86) .. (338.79,181.21) .. controls (338.79,183.57) and (337.61,184.75) .. (335.25,184.75) .. controls (332.9,184.75) and (331.72,185.93) .. (331.72,188.28) -- (330,190) -- (330,190) ;
\draw    (360,140) .. controls (357.64,140) and (356.46,138.82) .. (356.46,136.46) .. controls (356.46,134.11) and (355.28,132.93) .. (352.93,132.93) .. controls (350.57,132.93) and (349.39,131.75) .. (349.39,129.39) .. controls (349.39,127.04) and (348.21,125.86) .. (345.86,125.86) .. controls (343.5,125.86) and (342.32,124.68) .. (342.32,122.32) .. controls (342.32,119.97) and (341.14,118.79) .. (338.79,118.79) .. controls (336.43,118.79) and (335.25,117.61) .. (335.25,115.25) .. controls (335.25,112.9) and (334.07,111.72) .. (331.72,111.72) -- (330,110) -- (330,110) ;
\draw  [fill={rgb, 255:red, 128; green, 128; blue, 128 }  ,fill opacity=1 ] (350,150) .. controls (350,138.95) and (358.95,130) .. (370,130) .. controls (381.05,130) and (390,138.95) .. (390,150) .. controls (390,161.05) and (381.05,170) .. (370,170) .. controls (358.95,170) and (350,161.05) .. (350,150) -- cycle ;
\draw    (510,140) .. controls (510,137.64) and (511.18,136.46) .. (513.54,136.46) .. controls (515.89,136.46) and (517.07,135.28) .. (517.07,132.93) .. controls (517.07,130.57) and (518.25,129.39) .. (520.61,129.39) .. controls (522.96,129.39) and (524.14,128.21) .. (524.14,125.86) .. controls (524.14,123.5) and (525.32,122.32) .. (527.68,122.32) .. controls (530.03,122.32) and (531.21,121.14) .. (531.21,118.79) .. controls (531.21,116.43) and (532.39,115.25) .. (534.75,115.25) .. controls (537.1,115.25) and (538.28,114.07) .. (538.28,111.72) -- (540,110) -- (540,110) ;
\draw [line width=2.25]    (520,150) -- (550,150) ;
\draw  [fill={rgb, 255:red, 128; green, 128; blue, 128 }  ,fill opacity=1 ] (480,150) .. controls (480,138.95) and (488.95,130) .. (500,130) .. controls (511.05,130) and (520,138.95) .. (520,150) .. controls (520,161.05) and (511.05,170) .. (500,170) .. controls (488.95,170) and (480,161.05) .. (480,150) -- cycle ;
\draw  [line width=3]  (509,179.5) .. controls (509,179.22) and (509.22,179) .. (509.5,179) .. controls (509.78,179) and (510,179.22) .. (510,179.5) .. controls (510,179.78) and (509.78,180) .. (509.5,180) .. controls (509.22,180) and (509,179.78) .. (509,179.5) -- cycle ;
\draw  [line width=3]  (529,159.5) .. controls (529,159.22) and (529.22,159) .. (529.5,159) .. controls (529.78,159) and (530,159.22) .. (530,159.5) .. controls (530,159.78) and (529.78,160) .. (529.5,160) .. controls (529.22,160) and (529,159.78) .. (529,159.5) -- cycle ;
\draw  [line width=3]  (519,170.5) .. controls (519,170.22) and (519.22,170) .. (519.5,170) .. controls (519.78,170) and (520,170.22) .. (520,170.5) .. controls (520,170.78) and (519.78,171) .. (519.5,171) .. controls (519.22,171) and (519,170.78) .. (519,170.5) -- cycle ;

\draw (97,82.4) node [anchor=north west][inner sep=0.75pt]    {$n^{-2}$};
\draw (51,92.4) node [anchor=north west][inner sep=0.75pt]    {$1^{-}$};
\draw (31,141.4) node [anchor=north west][inner sep=0.75pt]    {$\hat{2}^{+}$};
\draw (51,192.4) node [anchor=north west][inner sep=0.75pt]    {$\hat{3}^{+}$};
\draw (151,142.4) node [anchor=north west][inner sep=0.75pt]    {$n-1^{+}$};
\draw (227,122.4) node [anchor=north west][inner sep=0.75pt]    {$\hat{k}_{2,3} \ on\ shell$};
\draw (421,120.4) node [anchor=north west][inner sep=0.75pt]    {$\hat{k}_{3,4} \ $};
\draw (487,75.4) node [anchor=north west][inner sep=0.75pt]    {$\hat{2}^{+}$};
\draw (327,86.4) node [anchor=north west][inner sep=0.75pt]    {$\hat{3}^{+}$};
\draw (327,192.4) node [anchor=north west][inner sep=0.75pt]    {$4^{+}$};
\draw (487,201.4) node [anchor=north west][inner sep=0.75pt]    {$5^{+}$};
\draw (540,91.4) node [anchor=north west][inner sep=0.75pt]    {$1^{-}$};
\draw (553,141.4) node [anchor=north west][inner sep=0.75pt]    {$n^{-2}$};
\draw (466,152.4) node [anchor=north west][inner sep=0.75pt]    {$+$};
\draw (392,153.4) node [anchor=north west][inner sep=0.75pt]    {$-$};

\end{tikzpicture}

other poles either have all plus helicity for three-point Yang-Mills amplitude or have single minus helicity for four or higher-point Yang-Mills amplitude. Which vanishes due to the mostly helicity violation (MHV) requirement.

\par The three-point pure Yang-Mills amplitude can be written as:
\begin{equation}
    \mathcal{A}(3^{+},4^{+},-k_{I}^{-})=\frac{\sbraket{34}^{4}}{\sbraket{34}\sbraket{4k_{I}}\sbraket{k_{I}3}}
\end{equation}	
where $I={3,4}$ and $k_{I}=k_{3,4}$, we construct the $n$ point amplitude from $n-1$ point amplitude using the three-point pure Yang-Mills amplitude as a building block:
\begin{equation}
\begin{aligned}
    &\mathcal{A}(1^{-},\hat{2}^{+},\hat{3}^{+}\dots,n-2^{+},n-1^{+}|\underline{\hat{n}}^{-2})
    \\
    =&\frac{\sbraket{ab}^{2}}{2m^{2}}\frac{\braket{1b}^{4}}{\braket{1\hat{2}}\braket{\hat{2}\hat{k}_{I}}\braket{\hat{k}_{I}5}\cdots\braket{n-1,1}} \times\frac{1}{\braket{34}\sbraket{34}}\times\frac{\sbraket{\hat{3}4}^{4}}{\sbraket{\hat{3}4}\sbraket{4\hat{k}_{I}}\sbraket{\hat{k}_{I}\hat{3}}},
\end{aligned}
\end{equation}
we can simplify the above equation using:
\begin{equation}
\begin{aligned}
    \sbraket{\hat{3}\hat{k}_{I}} \braket{\hat{k}_{I}5}&=\sbraket{34}\braket{45}
    \\
    \braket{\hat{2}\hat{k}_{I}} \sbraket{\hat{k}_{I}\hat{4}}&=\braket{23}\sbraket{34}.
\end{aligned}
\end{equation}
Finally, we get the $n$ point amplitude:
\begin{equation}\label{5.30}
    \mathcal{A}(1^{-},2^{+},\dots,n-2^{+},n-1^{+}|\underline{n}^{-2})=\frac{\sbraket{ab}^{2}}{2m^{2}}\frac{\braket{1b}^{4}}{\braket{12}\braket{23}\cdots\braket{n-1,1}}
\end{equation}
where the boundary comes from the $z$ dependence of the amplitude. More specifically, the large $z$ limit of the amplitude corresponds to the boundary. Since we shift momentum $2,3$, the naive $z$ dependence we can read from the amplitude exists in $\braket{23}\braket{34}$. Thus, the naive  $z$ dependence is $\frac{1}{z^{2}}$, which leads to the vanishing boundary. One can produce the n-point generalization of \eqref{differentspin}:
\begin{equation}
\begin{aligned}
    \mathcal{A}(1^{-},2^{+},\dots,n-2^{+},n-1^{+}|\underline{n}^{-2})&=\frac{\sbraket{ab}^{2}}{2m^{2}}\frac{\braket{1b}^{4}}{\braket{12}\braket{23}\cdots\braket{n-1,1}},
    \\
    \mathcal{A}(1^{-},2^{+},\dots,n-2^{+},n-1^{+}|\underline{n}^{-1})&=\frac{\sbraket{ab}^{2}}{m^{2}}\frac{\braket{1b}^{3}\braket{1a}}{\braket{12}\braket{23}\cdots\braket{n-1,1}},
    \\
    \mathcal{A}(1^{-},2^{+},\dots,n-2^{+},n-1^{+}|\underline{n}^{0})&=\frac{\sqrt{6}\sbraket{ab}^{2}}{2m^{2}} \frac{\braket{1b}^{2}\braket{1a}^{2}}{\braket{12}\braket{23}\cdots\braket{n-1,1}},
    \\
    \mathcal{A}(1^{-},2^{+},\dots,n-2^{+},n-1^{+}|\underline{n}^{+1})&=\frac{\sbraket{ab}^{2}}{m^{2}} \frac{\braket{1b}\braket{1a}^{3}}{\braket{12}\braket{23}\cdots\braket{n-1,1}},
    \\
    \mathcal{A}(1^{-},2^{+},\dots,n-2^{+},n-1^{+}|\underline{n}^{+2})&=\frac{\sbraket{ab}^{2}}{2m^{2}} \frac{\braket{1a}^{4}}{\braket{12}\braket{23}\cdots\braket{n-1,1}},
\end{aligned}
\end{equation}
using the raising and lowering operator as mentioned before
\section{Conclusion}
\par In this paper, we reviewed connections between tree-level amplitudes of twisted heterotic strings and conventional type-I superstrings \cite{Guillen:2021mwp}. By using B-G currents and their recursion, we conjectured a compact formula \eqref{formula} for the QFT building block of $n$-pt conventional type-I superstring amplitude of the spin-$2$ state at the first mass level coupled to $n-1$ gluons in spinor helicity basis. This formula can be regarded as a $n$ point generalization from the $4$-pt coupling of one single spin-$2$ massive state and massless states in \cite{Feng:2010yx}.  More importantly, by the choice of gluon helicities $(1^{-},2^{+},3^{+},\dots,n-1^{+})$, the $n$ point conjecture \eqref{formula} can be viewed as a massive extension of the Parker-Taylor formula \cite{Parke:1986gb} for pure-gluon tree amplitudes with MHV helicities. We then offered various pieces of evidence: We provided an analytic proof at $5$-pt and a numerical check at $6$-pt. We also applied the BCFW recursion to the QFT building block and found the expected result. 
\par One possible generalization of this topic is towards amplitudes with more massive states or higher excited states, one can find detailed studies for mass level 2 states in \cite{Bianchi:2010es, Feng:2011qc}. What's more, if we generalize to two higher excited massive states, such higher-spin amplitudes were recently used to study classical Kerr black-hole scattering \cite{Guevara:2018wpp, Chiodaroli:2021eug}. 
\par Another direct generalization is to consider the NMHV-liked QFT building block of conventional type-I string amplitude, which is having two or more massless particles having minus helicity. We have proven that the QFT building block with all massless particles having plus helicity vanishes, this is similar to the behavior of MHV violated Yang-Mills amplitude. One can expect, that the NMHV-liked amplitude would be more complicated, as how NMHV amplitude behaves in Yang-Mills theory. 
\section*{Acknowledgement}
I'm grateful to Oliver Schlotterer for inspiring discussion and related work. Moreover, Shruti Paranjape and Yi-Xiao Tao are thanked for their helpful discussion and valuable comments on the draft.
\appendix
\section{Proof of vanishing all-plus helicity configuration}\label{vanish}
\par In this appendix, we will discuss the QFT building block with all-plus helicity configuration\footnote{Still, the other spin choice of massive particle $n$ are generated by using raising and lowering operator as we did in \eqref{2.25}.}: $\mathcal{A}(1^{+},2^{+},\dots,n-2^{+},n-1^{+}|\underline{n}^{-2})$. We claim that the QFT building block with this helicity choice equals to zero.
\par We can't use the old reference spinor helicity choice as before since all the helicities of gluons are the same. In this appendix, we choose all the reference spinors of gluons to be $\bra{b}$, the polarization vectors can be written as:
\begin{equation}
    \left.\epsilon^{\mu}_{i} \right|_{r^{\mu}=b^{\mu}}=\frac{\bra{b}\bar{\sigma}^{\mu}\sket{i}}{\braket{bi}}
\end{equation}
\par By using B-G currents \eqref{amplitude} and its definition \eqref{currentsfieldstrengthandgauge}, we can express the QFT building block as we did in \eqref{5pointamplitude}, but with $P=1,2,3,\dots,n-1$this time:
\begin{equation}\label{allplus}
    \mathcal{A}(1^{+},2^{+},\dots,n-2^{+},n-1^{+}|\underline{n})= (\Phi_{P})^{\mu\nu}(\Phi_{n})_{\mu\nu}=\sqrt{\ap} \sum_{P=QR}  F_{Q \, \rho}^{\mu}F_{R}^{\rho\nu}(\Phi_{n})_{\mu\nu}  +{\rm cyc}_P.
\end{equation}
where we can apply \eqref{2.31} and expand $F_{Q \, \rho}^{\mu}$ and $F_{R}^{\rho\nu}$. Since we are considering the all plus helicity configuration, we can apply \eqref{shform} to $\epsilon_{X}^{\mu}$, where $X$ can be any multiple-particle label:
\begin{equation}
    \left.\epsilon^{\mu}(i^{+},i+1^{+},\dots ,n^{+})\right|_{r_{i,i+1,\dots,n}^{\mu}=b^{^{\mu}}}=\frac{\bra{b}\bar{\sigma}^{\mu}\slashed{k}_{i,i+1,\dots,n}\ket{b}}{\sqrt{2}\braket{bi}\braket{i,i+1}\cdots\braket{n-1,n}\braket{nb}}
\end{equation}
by using Fierz identity \eqref{Fierz}, we can easily find any contraction among polarization vectors or B-G currents equals to zero. For the same reasons, reference spinors $\ket{r_{i}}=\ket{b}$ lead to vanishing contractions $\epsilon^{\mu}_{P}(\Phi_{n})_{\mu\nu}$ for arbitrary multiparticle $P$. As a consequence of our choice of reference spinors,
\begin{equation}
    \left.\epsilon_{P}\cdot\epsilon_{Q} \right|_{r^{\mu}_{i}= b^{\mu}}=0,\qquad \left.\epsilon^{\mu}_{P}(\Phi_{n})_{\mu\nu}\right|_{r^{\mu}_{i}= b^{\mu}}=0,
\end{equation} 
each contribution $F_{Q \, \rho}^{\mu}F_{R}^{\rho\nu} (\Phi_{n})_{\mu\nu}$ to the B-G formula \eqref{amplitude} vanishes in the all-plus helicity configuration:
\begin{equation}\label{a.4}
\begin{aligned}
    &F_{Q \, \rho}^{\mu}F_{R}^{\rho\nu} \left.(\Phi_{n})_{\mu\nu}\right|_{r^{\mu}_{i}=b^{\mu}}
    \\
    =&\Big[k^{\mu}_{Q}\epsilon_{Q\rho} -k_{Q\rho} \epsilon^{\mu}_{Q} -\sum_{XY=Q}(\epsilon^{\mu}_{X}\epsilon_{Y\rho}-\epsilon_{X\rho}\epsilon^{\mu}_{Y})\Big]
    \\
    &\times\Big[k^{\rho}_{R}\epsilon_{R}^{\nu} -k_{R}^{\nu} \epsilon^{\rho}_{R} -\sum_{ZW=R}(\epsilon^{\rho}_{Z}\epsilon_{W}^{\nu}-\epsilon_{Z}^{\nu}\epsilon^{\rho}_{W})\Big]\left.(\Phi_{n})_{\mu\nu}\right|_{r^{\mu}_{i}=b^{\mu}}
    \\
    =&-k^{\mu}_{Q} k_{R}^{\nu} \left(\epsilon_{Q}\cdot\epsilon_{R}\right) \left.(\Phi_{n})_{\mu\nu}\right|_{r^{\mu}_{i}=b^{\mu}}=0
\end{aligned}
\end{equation}
where the second step follows from discarding any $\epsilon^{\mu}_{P}(\Phi_{n})_{\mu\nu}$, and we finally use the vanishing of $\epsilon_{Q}\cdot\epsilon_{R}$ in the last step. Thus, the QFT building block with all plus helicity choice vanishes.
\par Or we can use a similar power counting as in section \ref{4dim}, there are $n-1$ momentum vectors in each term, but there exist $n-1$ polarization vectors and one polarization tensor, which has $2$ indices. There are $n+1$ indices from polarizations in total. Thus, there exists at least one contraction between two Lorentz indices of the polarizations in each term. We can conclude that the QFT building block with all plus helicity choice vanishes.
\section{BCFW recursion with 1 and n shifted}
We can generalize the discussion to two or even more massive states in the future. Thus knowing how to apply the shift with massive particles to the QFT building block is useful, although we have tentative boundaries in this shift choice.
\subsection{BCFW shift for massive momentum}
\par The BCFW shift for all massless external momentum is already discussed in the review section. For the amplitude with massive legs, we need to shift the massive momentum $k_{j}$ and the massless momentum $k_{i}$ as follows \cite{Ballav:2020ese}:
	\begin{equation}
		\begin{aligned}
			\hat{k}^{\mu}_{j}&=k^{\mu}_{j}+zr^{\mu}
			\\
			\hat{k}^{\mu}_{i}&=k^{\mu}_{i}-zr^{\mu}
		\end{aligned}
	\end{equation}
	the momentum conservation is naturally satisfied. We expand ${k}^{\mu}_{j}$ into ${k}^{\mu}_{j}=a^{\mu}+b^{\mu}$.
	\par The on-shell condition becomes:
	\begin{equation}
		\begin{aligned}
			\hat{k}_{j}^{2}+m^{2}=0
			\\
			\hat{k}_{i}^{2}=0
		\end{aligned}
	\end{equation}
	\par Next, we choose a suitable $z$, which makes the complex shifted propagator on-shell. We can find the $z$ we want by solving the mass shell equation:
	\begin{equation}
		\begin{aligned}
			\hat{k}^{2}_{I}+m^{2}_{I}=(\sum_{i\in I}\hat{k}_{i})^{2}+m^{2}_{I}=0
			\\
			\hat{k}^{2}_{J}+m^{2}_{J}=(\sum_{j\in J}\hat{k}_{j})^{2}+m^{2}_{J}=0,
		\end{aligned}
	\end{equation}
	where $m^{2}_{I}$ and $m^{2}_{J}$ denotes the mass of $k^{\mu}_{I}$ and $k^{\mu}_{J}$. Since the only two shifted momentum is $k^{\mu}_{j}$ and $k^{\mu}_{i}$, the complex shift of $k^{\mu}_{I}$ and $k^{\mu}_{J}$ is also $z r^{\mu}$. Only single first-order singularity contribute residual. We need to set the second-order and higher-order to zero, which means:
	\begin{equation}
		r^{2}=0.
	\end{equation}
	The first constraint on $r^{\mu}$ is: $r^{\mu}$ has to be a null vector. Plugging back to the on-shell condition of momentum $1$ and $n$, we have:
	\begin{equation}
		k_{i}\cdot r=k_{j}\cdot r=0
	\end{equation}
	The second constrain on $r^{\mu}$ is: $r^{\mu}$ has to be orthogonal to both $k^{\mu}_{i}$ and $k^{\mu}_{j}$
	\par Momentum conservation is automatically satisfied.
	The simplest nontrivial $r^{\mu}$ I found is:
	\begin{equation}\label{r}
		r^{\mu}=-\frac{1}{2m}\bra{i}_{\dot{\beta}}k^{\dot{\beta}\alpha}_{j}\sigma^{\mu}_{\alpha\dot{\alpha}}\ket{i}^{\dot{\alpha}},
	\end{equation}
	we can also define $\slashed{r}$ by contracting $r^{\mu}$ with $\sigma_{\mu}$:
	\begin{equation}
		\begin{aligned}
			\slashed{r}_{\alpha\dot{\beta}}=&r^{\mu}(\sigma^{\nu})_{\alpha\dot{\beta}}\eta_{\mu\nu}=-\frac{1}{2m}\bra{i}_{\dot{\gamma}}k_{j}^{\nu}\bar{\sigma}_{\nu}^{\dot{\gamma}\delta}\sigma^{\mu}_{\delta\dot{\delta}}\ket{i}^{\dot{\delta}}\sigma^{\rho}_{\alpha\dot{\beta}}\eta_{\mu\rho}=-\frac{1}{m}(\slashed{k}_{j}\ket{i})_{\alpha}\bra{i}_{\dot{\beta}}
			\\
			=&\frac{1}{m}(\sket{a}_{\alpha}\braket{ai}+\sket{b}_{\alpha}\braket{bi})\bra{i}_{\dot{\beta}}.
		\end{aligned}
	\end{equation}
	\subsection{Application to the QFT building block}
	We can now apply our new BCFW recursion with massive external legs shifted into the QFT building block.
	\par The momentum is shifted as:
	\begin{equation}
		\hat{k}^{\mu}_{1}=k^{\mu}_{1}+z r^{\mu} \qquad\hat{k}^{\mu}_{n}=k^{\mu}_{n}-z r^{\mu}
	\end{equation}
	Where we still expand the massive leg into $a$ and $b$:
	\begin{equation}
		k^{\mu}_{n}=a^{\mu}+b^{\mu}
	\end{equation}
	where $n$ is the only massive leg. Plugging back to \eqref{r}, we get:
	\begin{equation}
		r^{\mu}=-\frac{1}{2m}\bra{1}_{\dot{\beta}}k^{\dot{\beta}\alpha}_{n}\sigma^{\mu}_{\alpha\dot{\alpha}}\ket{i}^{\dot{\alpha}},
	\end{equation}
	the $\slashed{r}$ is:
	\begin{equation}
		\begin{aligned}
			\slashed{r}_{\alpha\dot{\beta}}=&r^{\mu}(\sigma^{\nu})_{\alpha\dot{\beta}}\eta_{\mu\nu}=-\frac{1}{2m}\bra{1}_{\dot{\gamma}}k_{n}^{\nu}\bar{\sigma}_{\nu}^{\dot{\gamma}\delta}\sigma^{\mu}_{\delta\dot{\delta}}\ket{1}^{\dot{\delta}}\sigma^{\rho}_{\alpha\dot{\beta}}\eta_{\mu\rho}
			\\
			=&-\frac{1}{m}(\slashed{k}_{n}\ket{1})_{\alpha}\bra{1}_{\dot{\beta}}=\frac{1}{m}(\sket{a}_{\alpha}\braket{a1}+\sket{b}_{\alpha}\braket{b1})\bra{1}_{\dot{\beta}}.
		\end{aligned}
	\end{equation}
	\par As for the shifted spinor helicity, we have:
	\begin{equation}
		\begin{aligned}
			\hat{\slashed{k}}_{1}=\slashed{k}_{1}+z \slashed{r}&=\sket{1}\bra{1}+\frac{z}{m}(\sket{a}\braket{a1}+\sket{b}\braket{b1})\bra{1}
			\\
			\hat{\slashed{k}}_{n}=\slashed{a}+\slashed{b}-z \slashed{r}&=\sket{a}\bra{a}+\sket{b}\bra{b}-\frac{z}{m}(\sket{a}\braket{a1}+\sket{b}\braket{b1})\bra{1}
		\end{aligned}
	\end{equation}
	where momentum $p_{n}$ is decomposed into $p_{n}=a+b$. Thus, we have:
	\begin{equation}
		\begin{aligned}
			\sket{\hat{1}}&=\sket{1}+\frac{z}{m}(\sket{a}\braket{a1}+\sket{b}\braket{b1})
			\\
			\bra{\hat a}&=\bra{a}-\frac{z}{m}\braket{a1}\bra{1}
			\\
			\bra{\hat b}&=\bra{b}-\frac{z}{m}\braket{b1}\bra{1}
		\end{aligned}
	\end{equation}
	where other shifted spinor helicity is unchanged. We have several special relations:
	\begin{equation}
		\begin{aligned}
			\braket{\hat a\hat{1}}=\braket{a1}&\qquad\braket{\hat b\hat{1}}=\braket{b1}
			\\
			\sbraket{\hat a\hat{1}}=\sbraket{a1}&\qquad\sbraket{\hat b\hat{1}}=\sbraket{b1}
		\end{aligned}
	\end{equation}
	\par After rewriting four-point amplitude into spinor helicity form, we can get:
	\begin{equation}
		\mathcal{A}(1^{-},2^{+},3^{+}|\underline{4}^{-2})=\frac{\sbraket{ab}^{2}}{2m^{2}}\frac{\braket{1b}^{4}}{\braket{12}\braket{23}\braket{31}}
	\end{equation}
	where $k_{4}=a+b$, same as the discussion before for general $n$. 
	\par Suppose the $n-1$ point amplitudes takes the form:
	\begin{equation}
		\mathcal{A}(k_{I}^{-},3^{+},\dots,n-2^{+},n-1^{+}|\underline{n}^{-2})=\frac{\sbraket{ab}^{2}}{2m^{2}}\frac{\braket{k_{I}b}^{4}}{\braket{k_{I}3}\braket{34}\cdots\braket{n-1,k_{I}}}
	\end{equation}
	The three-point pure Yang-Mills amplitude can be written as:
	\begin{equation}
		\mathcal{A}(1^{-},2^{+},-k_{I}^{+})=\frac{\sbraket{2k_{I}}^{4}}{\sbraket{12}\sbraket{2k_{I}}\sbraket{k_{I}1}}
	\end{equation}	
where $I={1,2}$ and $k_{I}=k_{1,2}$, we can always construct the $n$ point amplitude from lower point amplitudes. Interestingly, there is still only one pole giving non-zero contribution:

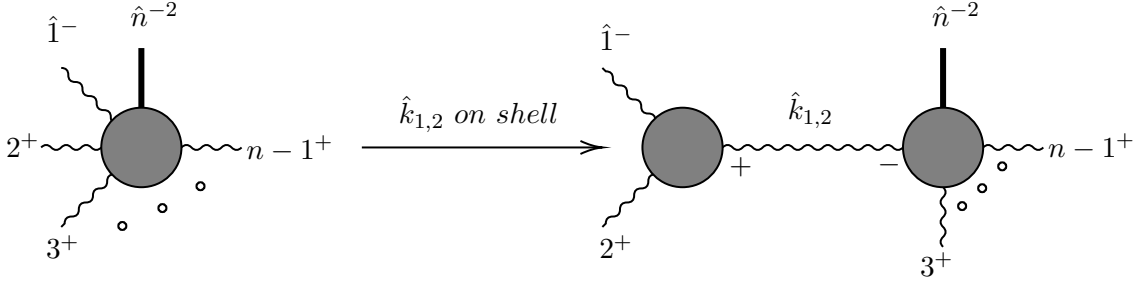
\tikzset{every picture/.style={line width=0.75pt}} 

\begin{tikzpicture}[x=0.75pt,y=0.75pt,yscale=-1,xscale=1]

\draw    (390,150) .. controls (391.67,148.33) and (393.33,148.33) .. (395,150) .. controls (396.67,151.67) and (398.33,151.67) .. (400,150) .. controls (401.67,148.33) and (403.33,148.33) .. (405,150) .. controls (406.67,151.67) and (408.33,151.67) .. (410,150) .. controls (411.67,148.33) and (413.33,148.33) .. (415,150) .. controls (416.67,151.67) and (418.33,151.67) .. (420,150) .. controls (421.67,148.33) and (423.33,148.33) .. (425,150) .. controls (426.67,151.67) and (428.33,151.67) .. (430,150) .. controls (431.67,148.33) and (433.33,148.33) .. (435,150) .. controls (436.67,151.67) and (438.33,151.67) .. (440,150) .. controls (441.67,148.33) and (443.33,148.33) .. (445,150) .. controls (446.67,151.67) and (448.33,151.67) .. (450,150) .. controls (451.67,148.33) and (453.33,148.33) .. (455,150) .. controls (456.67,151.67) and (458.33,151.67) .. (460,150) .. controls (461.67,148.33) and (463.33,148.33) .. (465,150) .. controls (466.67,151.67) and (468.33,151.67) .. (470,150) .. controls (471.67,148.33) and (473.33,148.33) .. (475,150) .. controls (476.67,151.67) and (478.33,151.67) .. (480,150) -- (480,150) ;
\draw    (550,150) .. controls (548.33,151.67) and (546.67,151.67) .. (545,150) .. controls (543.33,148.33) and (541.67,148.33) .. (540,150) .. controls (538.33,151.67) and (536.67,151.67) .. (535,150) .. controls (533.33,148.33) and (531.67,148.33) .. (530,150) .. controls (528.33,151.67) and (526.67,151.67) .. (525,150) .. controls (523.33,148.33) and (521.67,148.33) .. (520,150) -- (520,150) ;
\draw    (500,170) .. controls (501.67,171.67) and (501.67,173.33) .. (500,175) .. controls (498.33,176.67) and (498.33,178.33) .. (500,180) .. controls (501.67,181.67) and (501.67,183.33) .. (500,185) .. controls (498.33,186.67) and (498.33,188.33) .. (500,190) .. controls (501.67,191.67) and (501.67,193.33) .. (500,195) .. controls (498.33,196.67) and (498.33,198.33) .. (500,200) -- (500,200) ;
\draw    (120,150) .. controls (121.67,148.33) and (123.33,148.33) .. (125,150) .. controls (126.67,151.67) and (128.33,151.67) .. (130,150) .. controls (131.67,148.33) and (133.33,148.33) .. (135,150) .. controls (136.67,151.67) and (138.33,151.67) .. (140,150) .. controls (141.67,148.33) and (143.33,148.33) .. (145,150) .. controls (146.67,151.67) and (148.33,151.67) .. (150,150) -- (150,150) ;
\draw    (80,150) .. controls (78.33,151.67) and (76.67,151.67) .. (75,150) .. controls (73.33,148.33) and (71.67,148.33) .. (70,150) .. controls (68.33,151.67) and (66.67,151.67) .. (65,150) .. controls (63.33,148.33) and (61.67,148.33) .. (60,150) .. controls (58.33,151.67) and (56.67,151.67) .. (55,150) .. controls (53.33,148.33) and (51.67,148.33) .. (50,150) -- (50,150) ;
\draw    (60,110) .. controls (62.36,110) and (63.54,111.18) .. (63.54,113.54) .. controls (63.54,115.89) and (64.72,117.07) .. (67.07,117.07) .. controls (69.43,117.07) and (70.61,118.25) .. (70.61,120.61) .. controls (70.61,122.96) and (71.79,124.14) .. (74.14,124.14) .. controls (76.5,124.14) and (77.68,125.32) .. (77.68,127.68) .. controls (77.68,130.03) and (78.86,131.21) .. (81.21,131.21) .. controls (83.57,131.21) and (84.75,132.39) .. (84.75,134.75) .. controls (84.75,137.1) and (85.93,138.28) .. (88.28,138.28) -- (90,140) -- (90,140) ;
\draw    (90,160) .. controls (90,162.36) and (88.82,163.54) .. (86.46,163.54) .. controls (84.11,163.54) and (82.93,164.72) .. (82.93,167.07) .. controls (82.93,169.43) and (81.75,170.61) .. (79.39,170.61) .. controls (77.04,170.61) and (75.86,171.79) .. (75.86,174.14) .. controls (75.86,176.5) and (74.68,177.68) .. (72.32,177.68) .. controls (69.97,177.68) and (68.79,178.86) .. (68.79,181.21) .. controls (68.79,183.57) and (67.61,184.75) .. (65.25,184.75) .. controls (62.9,184.75) and (61.72,185.93) .. (61.72,188.28) -- (60,190) -- (60,190) ;
\draw  [fill={rgb, 255:red, 128; green, 128; blue, 128 }  ,fill opacity=1 ] (80,150) .. controls (80,138.95) and (88.95,130) .. (100,130) .. controls (111.05,130) and (120,138.95) .. (120,150) .. controls (120,161.05) and (111.05,170) .. (100,170) .. controls (88.95,170) and (80,161.05) .. (80,150) -- cycle ;
\draw [line width=2.25]    (100,130) -- (100,100) ;
\draw  [line width=3]  (90,189.5) .. controls (90,189.22) and (90.22,189) .. (90.5,189) .. controls (90.78,189) and (91,189.22) .. (91,189.5) .. controls (91,189.78) and (90.78,190) .. (90.5,190) .. controls (90.22,190) and (90,189.78) .. (90,189.5) -- cycle ;
\draw  [line width=3]  (110,180.5) .. controls (110,180.22) and (110.22,180) .. (110.5,180) .. controls (110.78,180) and (111,180.22) .. (111,180.5) .. controls (111,180.78) and (110.78,181) .. (110.5,181) .. controls (110.22,181) and (110,180.78) .. (110,180.5) -- cycle ;
\draw  [line width=3]  (129,169.5) .. controls (129,169.22) and (129.22,169) .. (129.5,169) .. controls (129.78,169) and (130,169.22) .. (130,169.5) .. controls (130,169.78) and (129.78,170) .. (129.5,170) .. controls (129.22,170) and (129,169.78) .. (129,169.5) -- cycle ;
\draw    (210,150) -- (328,150) ;
\draw [shift={(330,150)}, rotate = 180] [color={rgb, 255:red, 0; green, 0; blue, 0 }  ][line width=0.75]    (10.93,-3.29) .. controls (6.95,-1.4) and (3.31,-0.3) .. (0,0) .. controls (3.31,0.3) and (6.95,1.4) .. (10.93,3.29)   ;
\draw    (360,160) .. controls (360,162.36) and (358.82,163.54) .. (356.46,163.54) .. controls (354.11,163.54) and (352.93,164.72) .. (352.93,167.07) .. controls (352.93,169.43) and (351.75,170.61) .. (349.39,170.61) .. controls (347.04,170.61) and (345.86,171.79) .. (345.86,174.14) .. controls (345.86,176.5) and (344.68,177.68) .. (342.32,177.68) .. controls (339.97,177.68) and (338.79,178.86) .. (338.79,181.21) .. controls (338.79,183.57) and (337.61,184.75) .. (335.25,184.75) .. controls (332.9,184.75) and (331.72,185.93) .. (331.72,188.28) -- (330,190) -- (330,190) ;
\draw    (360,140) .. controls (357.64,140) and (356.46,138.82) .. (356.46,136.46) .. controls (356.46,134.11) and (355.28,132.93) .. (352.93,132.93) .. controls (350.57,132.93) and (349.39,131.75) .. (349.39,129.39) .. controls (349.39,127.04) and (348.21,125.86) .. (345.86,125.86) .. controls (343.5,125.86) and (342.32,124.68) .. (342.32,122.32) .. controls (342.32,119.97) and (341.14,118.79) .. (338.79,118.79) .. controls (336.43,118.79) and (335.25,117.61) .. (335.25,115.25) .. controls (335.25,112.9) and (334.07,111.72) .. (331.72,111.72) -- (330,110) -- (330,110) ;
\draw  [fill={rgb, 255:red, 128; green, 128; blue, 128 }  ,fill opacity=1 ] (350,150) .. controls (350,138.95) and (358.95,130) .. (370,130) .. controls (381.05,130) and (390,138.95) .. (390,150) .. controls (390,161.05) and (381.05,170) .. (370,170) .. controls (358.95,170) and (350,161.05) .. (350,150) -- cycle ;
\draw [line width=2.25]    (500,130) -- (500,100) ;
\draw  [fill={rgb, 255:red, 128; green, 128; blue, 128 }  ,fill opacity=1 ] (480,150) .. controls (480,138.95) and (488.95,130) .. (500,130) .. controls (511.05,130) and (520,138.95) .. (520,150) .. controls (520,161.05) and (511.05,170) .. (500,170) .. controls (488.95,170) and (480,161.05) .. (480,150) -- cycle ;
\draw  [line width=3]  (509,179.5) .. controls (509,179.22) and (509.22,179) .. (509.5,179) .. controls (509.78,179) and (510,179.22) .. (510,179.5) .. controls (510,179.78) and (509.78,180) .. (509.5,180) .. controls (509.22,180) and (509,179.78) .. (509,179.5) -- cycle ;
\draw  [line width=3]  (529,159.5) .. controls (529,159.22) and (529.22,159) .. (529.5,159) .. controls (529.78,159) and (530,159.22) .. (530,159.5) .. controls (530,159.78) and (529.78,160) .. (529.5,160) .. controls (529.22,160) and (529,159.78) .. (529,159.5) -- cycle ;
\draw  [line width=3]  (519,170.5) .. controls (519,170.22) and (519.22,170) .. (519.5,170) .. controls (519.78,170) and (520,170.22) .. (520,170.5) .. controls (520,170.78) and (519.78,171) .. (519.5,171) .. controls (519.22,171) and (519,170.78) .. (519,170.5) -- cycle ;

\draw (93,75.4) node [anchor=north west][inner sep=0.75pt]    {$\hat{n}^{-2}$};
\draw (51,82.4) node [anchor=north west][inner sep=0.75pt]    {$\hat{1}^{-}$};
\draw (31,141.4) node [anchor=north west][inner sep=0.75pt]    {$2^{+}$};
\draw (51,192.4) node [anchor=north west][inner sep=0.75pt]    {$3^{+}$};
\draw (151,142.4) node [anchor=north west][inner sep=0.75pt]    {$n-1^{+}$};
\draw (227,122.4) node [anchor=north west][inner sep=0.75pt]    {$\hat{k}_{1,2} \ on\ shell$};
\draw (421,120.4) node [anchor=north west][inner sep=0.75pt]    {$\hat{k}_{1,2} \ $};
\draw (551,141.4) node [anchor=north west][inner sep=0.75pt]    {$n-1^{+}$};
\draw (327,86.4) node [anchor=north west][inner sep=0.75pt]    {$\hat{1}^{-}$};
\draw (327,192.4) node [anchor=north west][inner sep=0.75pt]    {$2^{+}$};
\draw (487,201.4) node [anchor=north west][inner sep=0.75pt]    {$3^{+}$};
\draw (493,75.4) node [anchor=north west][inner sep=0.75pt]    {$\hat{n}^{-2}$};
\draw (466,152.4) node [anchor=north west][inner sep=0.75pt]    {$-$};
\draw (392,153.4) node [anchor=north west][inner sep=0.75pt]    {$+$};
\end{tikzpicture}
the helicity choice on the propagator has to be plus on the left and minus to the right because we have shown that the all-plus configuration vanishes. And we can only have a three-point Yang-Mills amplitude on the left, due to the MHV.
\begin{equation}
\begin{aligned}
    &\mathcal{A}(\hat{1}^{-},2^{+},\dots,n-2^{+},n-1^{+}|\underline{\hat{n}}^{-2})
    \\
    =&\frac{\sbraket{ab}^{2}}{2m^{2}}\frac{\braket{\hat{k}_{I}\hat b}^{4}}{\braket{\hat{k}_{I}3}\braket{34}\cdots\braket{n-1,\hat{k}_{I}}} \times\frac{1}{\braket{\hat{1}2}\sbraket{\hat{1}2}}\times\frac{\sbraket{2\hat{k}_{I}}^{4}}{\sbraket{\hat{1}2}\sbraket{2\hat{k}_{I}}\sbraket{\hat{k}_{I}\hat{1}}}
\end{aligned}
\end{equation}
where we ignore the boundary term for now. We can simplify the above equation using the following:
\begin{equation}
\begin{aligned}
    \sbraket{2\hat{k}_{I}} \braket{\hat{k}_{I}\hat{b}}&=-\sbra{2}\hat{1}\ket{\hat a}=\sbraket{2\hat{1}}\braket{1b}
    \\
    \sbraket{\hat{1}\hat{k}_{I}} \braket{\hat{k}_{I}3}&=\sbraket{\hat{1}2}\braket{23}
    \\
    \braket{n-1,\hat{k}_{I}}\sbraket{\hat{k}_{I}2}&=\braket{n-1,\hat{1}}\sbraket{\hat{1}2}.
\end{aligned}
\end{equation}
\par Finally, we get the $n$ point amplitude:
\begin{equation}\label{5.20}
    \mathcal{A}(\hat{1}^{-},2^{+},\dots,n-2^{+},n-1^{+}|\underline{\hat{n}}^{-2})=\frac{\sbraket{ab}^{2}}{2m^{2}}\frac{\braket{1b}^{4}}{\braket{12}\braket{23}\cdots\braket{n-1,1}}
\end{equation}
this is what we expected for $n$ point amplitude. It agrees with the nonzero configuration we analytically worked out in section \ref{npt}.
\par However, there is one more thing noticeable, we also need to consider the boundary contribution. we can find the naive boundary in \eqref{5.20}
\bibliography{biblio}

\begin{thebibliography}{41}%
\makeatletter
\providecommand \@ifxundefined [1]{%
 \@ifx{#1\undefined}
}%
\providecommand \@ifnum [1]{%
 \ifnum #1\expandafter \@firstoftwo
 \else \expandafter \@secondoftwo
 \fi
}%
\providecommand \@ifx [1]{%
 \ifx #1\expandafter \@firstoftwo
 \else \expandafter \@secondoftwo
 \fi
}%
\providecommand \natexlab [1]{#1}%
\providecommand \enquote  [1]{``#1''}%
\providecommand \bibnamefont  [1]{#1}%
\providecommand \bibfnamefont [1]{#1}%
\providecommand \citenamefont [1]{#1}%
\providecommand \href@noop [0]{\@secondoftwo}%
\providecommand \href [0]{\begingroup \@sanitize@url \@href}%
\providecommand \@href[1]{\@@startlink{#1}\@@href}%
\providecommand \@@href[1]{\endgroup#1\@@endlink}%
\providecommand \@sanitize@url [0]{\catcode `\\12\catcode `\$12\catcode
  `\&12\catcode `\#12\catcode `\^12\catcode `\_12\catcode `\%12\relax}%
\providecommand \@@startlink[1]{}%
\providecommand \@@endlink[0]{}%
\providecommand \url  [0]{\begingroup\@sanitize@url \@url }%
\providecommand \@url [1]{\endgroup\@href {#1}{\urlprefix }}%
\providecommand \urlprefix  [0]{URL }%
\providecommand \Eprint [0]{\href }%
\providecommand \doibase [0]{http://dx.doi.org/}%
\providecommand \selectlanguage [0]{\@gobble}%
\providecommand \bibinfo  [0]{\@secondoftwo}%
\providecommand \bibfield  [0]{\@secondoftwo}%
\providecommand \translation [1]{[#1]}%
\providecommand \BibitemOpen [0]{}%
\providecommand \bibitemStop [0]{}%
\providecommand \bibitemNoStop [0]{.\EOS\space}%
\providecommand \EOS [0]{\spacefactor3000\relax}%
\providecommand \BibitemShut  [1]{\csname bibitem#1\endcsname}%
\let\auto@bib@innerbib\@empty
\bibitem [{\citenamefont {Veneziano}(1968)}]{Veneziano:1968yb}%
  \BibitemOpen
  \bibfield  {author} {\bibinfo {author} {\bibfnamefont {G.}~\bibnamefont
  {Veneziano}},\ }\href {\doibase 10.1007/BF02824451} {\bibfield  {journal}
  {\bibinfo  {journal} {Nuovo Cim. A}\ }\textbf {\bibinfo {volume} {57}},\
  \bibinfo {pages} {190} (\bibinfo {year} {1968})}\BibitemShut {NoStop}%
\bibitem [{\citenamefont {Kawai}\ \emph {et~al.}(1986)\citenamefont {Kawai},
  \citenamefont {Lewellen},\ and\ \citenamefont {Tye}}]{Kawai:1985xq}%
  \BibitemOpen
  \bibfield  {author} {\bibinfo {author} {\bibfnamefont {H.}~\bibnamefont
  {Kawai}}, \bibinfo {author} {\bibfnamefont {D.~C.}\ \bibnamefont {Lewellen}},
  \ and\ \bibinfo {author} {\bibfnamefont {S.~H.~H.}\ \bibnamefont {Tye}},\
  }\href {\doibase 10.1016/0550-3213(86)90362-7} {\bibfield  {journal}
  {\bibinfo  {journal} {Nucl. Phys. B}\ }\textbf {\bibinfo {volume} {269}},\
  \bibinfo {pages} {1} (\bibinfo {year} {1986})}\BibitemShut {NoStop}%
\bibitem [{\citenamefont {Bern}\ \emph {et~al.}(2008)\citenamefont {Bern},
  \citenamefont {Carrasco},\ and\ \citenamefont {Johansson}}]{Bern:2008qj}%
  \BibitemOpen
  \bibfield  {author} {\bibinfo {author} {\bibfnamefont {Z.}~\bibnamefont
  {Bern}}, \bibinfo {author} {\bibfnamefont {J.~J.~M.}\ \bibnamefont
  {Carrasco}}, \ and\ \bibinfo {author} {\bibfnamefont {H.}~\bibnamefont
  {Johansson}},\ }\href {\doibase 10.1103/PhysRevD.78.085011} {\bibfield
  {journal} {\bibinfo  {journal} {Phys. Rev. D}\ }\textbf {\bibinfo {volume}
  {78}},\ \bibinfo {pages} {085011} (\bibinfo {year} {2008})},\ \Eprint
  {http://arxiv.org/abs/0805.3993} {arXiv:0805.3993 [hep-ph]} \BibitemShut
  {NoStop}%
\bibitem [{\citenamefont {Bern}\ \emph {et~al.}(2019)\citenamefont {Bern},
  \citenamefont {Carrasco}, \citenamefont {Chiodaroli}, \citenamefont
  {Johansson},\ and\ \citenamefont {Roiban}}]{Bern:2019prr}%
  \BibitemOpen
  \bibfield  {author} {\bibinfo {author} {\bibfnamefont {Z.}~\bibnamefont
  {Bern}}, \bibinfo {author} {\bibfnamefont {J.~J.}\ \bibnamefont {Carrasco}},
  \bibinfo {author} {\bibfnamefont {M.}~\bibnamefont {Chiodaroli}}, \bibinfo
  {author} {\bibfnamefont {H.}~\bibnamefont {Johansson}}, \ and\ \bibinfo
  {author} {\bibfnamefont {R.}~\bibnamefont {Roiban}},\ }\href@noop {} {\
  (\bibinfo {year} {2019})},\ \Eprint {http://arxiv.org/abs/1909.01358}
  {arXiv:1909.01358 [hep-th]} \BibitemShut {NoStop}%
\bibitem [{\citenamefont {D'Hoker}\ and\ \citenamefont
  {Phong}(1989)}]{DHoker:1989cxq}%
  \BibitemOpen
  \bibfield  {author} {\bibinfo {author} {\bibfnamefont {E.}~\bibnamefont
  {D'Hoker}}\ and\ \bibinfo {author} {\bibfnamefont {D.~H.}\ \bibnamefont
  {Phong}},\ }\href {\doibase 10.1007/BF01218413} {\bibfield  {journal}
  {\bibinfo  {journal} {Commun. Math. Phys.}\ }\textbf {\bibinfo {volume}
  {125}},\ \bibinfo {pages} {469} (\bibinfo {year} {1989})}\BibitemShut
  {NoStop}%
\bibitem [{\citenamefont {Bern}\ \emph {et~al.}(2010)\citenamefont {Bern},
  \citenamefont {Carrasco},\ and\ \citenamefont {Johansson}}]{Bern:2010ue}%
  \BibitemOpen
  \bibfield  {author} {\bibinfo {author} {\bibfnamefont {Z.}~\bibnamefont
  {Bern}}, \bibinfo {author} {\bibfnamefont {J.~J.~M.}\ \bibnamefont
  {Carrasco}}, \ and\ \bibinfo {author} {\bibfnamefont {H.}~\bibnamefont
  {Johansson}},\ }\href {\doibase 10.1103/PhysRevLett.105.061602} {\bibfield
  {journal} {\bibinfo  {journal} {Phys. Rev. Lett.}\ }\textbf {\bibinfo
  {volume} {105}},\ \bibinfo {pages} {061602} (\bibinfo {year} {2010})},\
  \Eprint {http://arxiv.org/abs/1004.0476} {arXiv:1004.0476 [hep-th]}
  \BibitemShut {NoStop}%
\bibitem [{\citenamefont {He}\ and\ \citenamefont
  {Schlotterer}(2017)}]{He:2016mzd}%
  \BibitemOpen
  \bibfield  {author} {\bibinfo {author} {\bibfnamefont {S.}~\bibnamefont
  {He}}\ and\ \bibinfo {author} {\bibfnamefont {O.}~\bibnamefont
  {Schlotterer}},\ }\href {\doibase 10.1103/PhysRevLett.118.161601} {\bibfield
  {journal} {\bibinfo  {journal} {Phys. Rev. Lett.}\ }\textbf {\bibinfo
  {volume} {118}},\ \bibinfo {pages} {161601} (\bibinfo {year} {2017})},\
  \Eprint {http://arxiv.org/abs/1612.00417} {arXiv:1612.00417 [hep-th]}
  \BibitemShut {NoStop}%
\bibitem [{\citenamefont {Stieberger}(2022)}]{Stieberger:2022lss}%
  \BibitemOpen
  \bibfield  {author} {\bibinfo {author} {\bibfnamefont {S.}~\bibnamefont
  {Stieberger}},\ }\href@noop {} {\  (\bibinfo {year} {2022})},\ \Eprint
  {http://arxiv.org/abs/2212.06816} {arXiv:2212.06816 [hep-th]} \BibitemShut
  {NoStop}%
\bibitem [{\citenamefont {Mafra}\ \emph {et~al.}(2013)\citenamefont {Mafra},
  \citenamefont {Schlotterer},\ and\ \citenamefont
  {Stieberger}}]{Mafra:2011nv}%
  \BibitemOpen
  \bibfield  {author} {\bibinfo {author} {\bibfnamefont {C.~R.}\ \bibnamefont
  {Mafra}}, \bibinfo {author} {\bibfnamefont {O.}~\bibnamefont {Schlotterer}},
  \ and\ \bibinfo {author} {\bibfnamefont {S.}~\bibnamefont {Stieberger}},\
  }\href {\doibase 10.1016/j.nuclphysb.2013.04.023} {\bibfield  {journal}
  {\bibinfo  {journal} {Nucl. Phys. B}\ }\textbf {\bibinfo {volume} {873}},\
  \bibinfo {pages} {419} (\bibinfo {year} {2013})},\ \Eprint
  {http://arxiv.org/abs/1106.2645} {arXiv:1106.2645 [hep-th]} \BibitemShut
  {NoStop}%
\bibitem [{\citenamefont {Broedel}\ \emph {et~al.}(2013)\citenamefont
  {Broedel}, \citenamefont {Schlotterer},\ and\ \citenamefont
  {Stieberger}}]{Broedel:2013tta}%
  \BibitemOpen
  \bibfield  {author} {\bibinfo {author} {\bibfnamefont {J.}~\bibnamefont
  {Broedel}}, \bibinfo {author} {\bibfnamefont {O.}~\bibnamefont
  {Schlotterer}}, \ and\ \bibinfo {author} {\bibfnamefont {S.}~\bibnamefont
  {Stieberger}},\ }\href {\doibase 10.1002/prop.201300019} {\bibfield
  {journal} {\bibinfo  {journal} {Fortsch. Phys.}\ }\textbf {\bibinfo {volume}
  {61}},\ \bibinfo {pages} {812} (\bibinfo {year} {2013})},\ \Eprint
  {http://arxiv.org/abs/1304.7267} {arXiv:1304.7267 [hep-th]} \BibitemShut
  {NoStop}%
\bibitem [{\citenamefont {Huang}\ \emph
  {et~al.}(2016{\natexlab{a}})\citenamefont {Huang}, \citenamefont
  {Schlotterer},\ and\ \citenamefont {Wen}}]{Huang:2016tag}%
  \BibitemOpen
  \bibfield  {author} {\bibinfo {author} {\bibfnamefont {Y.-t.}\ \bibnamefont
  {Huang}}, \bibinfo {author} {\bibfnamefont {O.}~\bibnamefont {Schlotterer}},
  \ and\ \bibinfo {author} {\bibfnamefont {C.}~\bibnamefont {Wen}},\ }\href
  {\doibase 10.1007/JHEP09(2016)155} {\bibfield  {journal} {\bibinfo  {journal}
  {JHEP}\ }\textbf {\bibinfo {volume} {09}},\ \bibinfo {pages} {155} (\bibinfo
  {year} {2016}{\natexlab{a}})},\ \Eprint {http://arxiv.org/abs/1602.01674}
  {arXiv:1602.01674 [hep-th]} \BibitemShut {NoStop}%
\bibitem [{\citenamefont {Azevedo}\ \emph {et~al.}(2018)\citenamefont
  {Azevedo}, \citenamefont {Chiodaroli}, \citenamefont {Johansson},\ and\
  \citenamefont {Schlotterer}}]{Azevedo:2018dgo}%
  \BibitemOpen
  \bibfield  {author} {\bibinfo {author} {\bibfnamefont {T.}~\bibnamefont
  {Azevedo}}, \bibinfo {author} {\bibfnamefont {M.}~\bibnamefont {Chiodaroli}},
  \bibinfo {author} {\bibfnamefont {H.}~\bibnamefont {Johansson}}, \ and\
  \bibinfo {author} {\bibfnamefont {O.}~\bibnamefont {Schlotterer}},\ }\href
  {\doibase 10.1007/JHEP10(2018)012} {\bibfield  {journal} {\bibinfo  {journal}
  {JHEP}\ }\textbf {\bibinfo {volume} {10}},\ \bibinfo {pages} {012} (\bibinfo
  {year} {2018})},\ \Eprint {http://arxiv.org/abs/1803.05452} {arXiv:1803.05452
  [hep-th]} \BibitemShut {NoStop}%
\bibitem [{\citenamefont {Guillen}\ \emph {et~al.}(2021)\citenamefont
  {Guillen}, \citenamefont {Johansson}, \citenamefont {Jusinskas},\ and\
  \citenamefont {Schlotterer}}]{Guillen:2021mwp}%
  \BibitemOpen
  \bibfield  {author} {\bibinfo {author} {\bibfnamefont {M.}~\bibnamefont
  {Guillen}}, \bibinfo {author} {\bibfnamefont {H.}~\bibnamefont {Johansson}},
  \bibinfo {author} {\bibfnamefont {R.~L.}\ \bibnamefont {Jusinskas}}, \ and\
  \bibinfo {author} {\bibfnamefont {O.}~\bibnamefont {Schlotterer}},\ }\href
  {\doibase 10.1103/PhysRevLett.127.051601} {\bibfield  {journal} {\bibinfo
  {journal} {Phys. Rev. Lett.}\ }\textbf {\bibinfo {volume} {127}},\ \bibinfo
  {pages} {051601} (\bibinfo {year} {2021})},\ \Eprint
  {http://arxiv.org/abs/2104.03314} {arXiv:2104.03314 [hep-th]} \BibitemShut
  {NoStop}%
\bibitem [{\citenamefont {Kashyap}\ \emph {et~al.}(2023)\citenamefont
  {Kashyap}, \citenamefont {Mafra}, \citenamefont {Verma},\ and\ \citenamefont
  {Ypanaqu\'e}}]{Kashyap:2023cdi}%
  \BibitemOpen
  \bibfield  {author} {\bibinfo {author} {\bibfnamefont {S.~P.}\ \bibnamefont
  {Kashyap}}, \bibinfo {author} {\bibfnamefont {C.~R.}\ \bibnamefont {Mafra}},
  \bibinfo {author} {\bibfnamefont {M.}~\bibnamefont {Verma}}, \ and\ \bibinfo
  {author} {\bibfnamefont {L.~A.}\ \bibnamefont {Ypanaqu\'e}},\ }\href@noop {}
  {\  (\bibinfo {year} {2023})},\ \Eprint {http://arxiv.org/abs/2311.12100}
  {arXiv:2311.12100 [hep-th]} \BibitemShut {NoStop}%
\bibitem [{\citenamefont {Chen~Huang}()}]{HuangShruti}%
  \BibitemOpen
  \bibfield  {author} {\bibinfo {author} {\bibfnamefont {S.~P.}\ \bibnamefont
  {Chen~Huang}},\ }\href@noop {} {\ }\Eprint {http://arxiv.org/abs/to appear}
  {to appear} \BibitemShut {NoStop}%
\bibitem [{\citenamefont {Parke}\ and\ \citenamefont
  {Taylor}(1986)}]{Parke:1986gb}%
  \BibitemOpen
  \bibfield  {author} {\bibinfo {author} {\bibfnamefont {S.~J.}\ \bibnamefont
  {Parke}}\ and\ \bibinfo {author} {\bibfnamefont {T.~R.}\ \bibnamefont
  {Taylor}},\ }\href {\doibase 10.1103/PhysRevLett.56.2459} {\bibfield
  {journal} {\bibinfo  {journal} {Phys. Rev. Lett.}\ }\textbf {\bibinfo
  {volume} {56}},\ \bibinfo {pages} {2459} (\bibinfo {year}
  {1986})}\BibitemShut {NoStop}%
\bibitem [{\citenamefont {Elvang}\ and\ \citenamefont
  {Huang}(2015)}]{Elvang:2015rqa}%
  \BibitemOpen
  \bibfield  {author} {\bibinfo {author} {\bibfnamefont {H.}~\bibnamefont
  {Elvang}}\ and\ \bibinfo {author} {\bibfnamefont {Y.-t.}\ \bibnamefont
  {Huang}},\ }\href@noop {} {\emph {\bibinfo {title} {{Scattering Amplitudes in
  Gauge Theory and Gravity}}}}\ (\bibinfo  {publisher} {Cambridge University
  Press},\ \bibinfo {year} {2015})\BibitemShut {NoStop}%
\bibitem [{\citenamefont {Britto}\ \emph
  {et~al.}(2005{\natexlab{a}})\citenamefont {Britto}, \citenamefont {Cachazo},
  \citenamefont {Feng},\ and\ \citenamefont {Witten}}]{Britto:2005fq}%
  \BibitemOpen
  \bibfield  {author} {\bibinfo {author} {\bibfnamefont {R.}~\bibnamefont
  {Britto}}, \bibinfo {author} {\bibfnamefont {F.}~\bibnamefont {Cachazo}},
  \bibinfo {author} {\bibfnamefont {B.}~\bibnamefont {Feng}}, \ and\ \bibinfo
  {author} {\bibfnamefont {E.}~\bibnamefont {Witten}},\ }\href {\doibase
  10.1103/PhysRevLett.94.181602} {\bibfield  {journal} {\bibinfo  {journal}
  {Phys. Rev. Lett.}\ }\textbf {\bibinfo {volume} {94}},\ \bibinfo {pages}
  {181602} (\bibinfo {year} {2005}{\natexlab{a}})},\ \Eprint
  {http://arxiv.org/abs/hep-th/0501052} {arXiv:hep-th/0501052} \BibitemShut
  {NoStop}%
\bibitem [{\citenamefont {Berends}\ and\ \citenamefont
  {Giele}(1988)}]{Berends:1987me}%
  \BibitemOpen
  \bibfield  {author} {\bibinfo {author} {\bibfnamefont {F.~A.}\ \bibnamefont
  {Berends}}\ and\ \bibinfo {author} {\bibfnamefont {W.~T.}\ \bibnamefont
  {Giele}},\ }\href {\doibase 10.1016/0550-3213(88)90442-7} {\bibfield
  {journal} {\bibinfo  {journal} {Nucl. Phys. B}\ }\textbf {\bibinfo {volume}
  {306}},\ \bibinfo {pages} {759} (\bibinfo {year} {1988})}\BibitemShut
  {NoStop}%
\bibitem [{\citenamefont {Berends}\ \emph {et~al.}(1988)\citenamefont
  {Berends}, \citenamefont {Giele},\ and\ \citenamefont
  {Kuijf}}]{Berends:1988zp}%
  \BibitemOpen
  \bibfield  {author} {\bibinfo {author} {\bibfnamefont {F.~A.}\ \bibnamefont
  {Berends}}, \bibinfo {author} {\bibfnamefont {W.~T.}\ \bibnamefont {Giele}},
  \ and\ \bibinfo {author} {\bibfnamefont {H.}~\bibnamefont {Kuijf}},\ }\href
  {\doibase 10.1016/0370-2693(88)90813-1} {\bibfield  {journal} {\bibinfo
  {journal} {Phys. Lett. B}\ }\textbf {\bibinfo {volume} {211}},\ \bibinfo
  {pages} {91} (\bibinfo {year} {1988})}\BibitemShut {NoStop}%
\bibitem [{\citenamefont {Berends}\ and\ \citenamefont
  {Giele}(1989)}]{Berends:1988zn}%
  \BibitemOpen
  \bibfield  {author} {\bibinfo {author} {\bibfnamefont {F.~A.}\ \bibnamefont
  {Berends}}\ and\ \bibinfo {author} {\bibfnamefont {W.~T.}\ \bibnamefont
  {Giele}},\ }\href {\doibase 10.1016/0550-3213(89)90398-2} {\bibfield
  {journal} {\bibinfo  {journal} {Nucl. Phys. B}\ }\textbf {\bibinfo {volume}
  {313}},\ \bibinfo {pages} {595} (\bibinfo {year} {1989})}\BibitemShut
  {NoStop}%
\bibitem [{\citenamefont {Tao}(2023{\natexlab{a}})}]{Tao:2023wls}%
  \BibitemOpen
  \bibfield  {author} {\bibinfo {author} {\bibfnamefont {Y.-X.}\ \bibnamefont
  {Tao}},\ }\href {\doibase 10.1007/JHEP09(2023)193} {\bibfield  {journal}
  {\bibinfo  {journal} {JHEP}\ }\textbf {\bibinfo {volume} {09}},\ \bibinfo
  {pages} {193} (\bibinfo {year} {2023}{\natexlab{a}})},\ \Eprint
  {http://arxiv.org/abs/2307.14772} {arXiv:2307.14772 [hep-th]} \BibitemShut
  {NoStop}%
\bibitem [{\citenamefont {Chen}\ and\ \citenamefont
  {Tao}(2023)}]{Chen:2023bji}%
  \BibitemOpen
  \bibfield  {author} {\bibinfo {author} {\bibfnamefont {Q.}~\bibnamefont
  {Chen}}\ and\ \bibinfo {author} {\bibfnamefont {Y.-X.}\ \bibnamefont {Tao}},\
  }\href {\doibase 10.1007/JHEP08(2023)038} {\bibfield  {journal} {\bibinfo
  {journal} {JHEP}\ }\textbf {\bibinfo {volume} {08}},\ \bibinfo {pages} {038}
  (\bibinfo {year} {2023})},\ \Eprint {http://arxiv.org/abs/2301.08043}
  {arXiv:2301.08043 [hep-th]} \BibitemShut {NoStop}%
\bibitem [{\citenamefont {Tao}(2023{\natexlab{b}})}]{Tao:2023yxy}%
  \BibitemOpen
  \bibfield  {author} {\bibinfo {author} {\bibfnamefont {Y.-X.}\ \bibnamefont
  {Tao}},\ }\href {\doibase 10.1103/PhysRevD.108.125020} {\bibfield  {journal}
  {\bibinfo  {journal} {Phys. Rev. D}\ }\textbf {\bibinfo {volume} {108}},\
  \bibinfo {pages} {125020} (\bibinfo {year} {2023}{\natexlab{b}})},\ \Eprint
  {http://arxiv.org/abs/2309.15657} {arXiv:2309.15657 [hep-th]} \BibitemShut
  {NoStop}%
\bibitem [{\citenamefont {Tao}\ and\ \citenamefont {Chen}(2023)}]{Tao:2022nqc}%
  \BibitemOpen
  \bibfield  {author} {\bibinfo {author} {\bibfnamefont {Y.-X.}\ \bibnamefont
  {Tao}}\ and\ \bibinfo {author} {\bibfnamefont {Q.}~\bibnamefont {Chen}},\
  }\href {\doibase 10.1007/JHEP02(2023)030} {\bibfield  {journal} {\bibinfo
  {journal} {JHEP}\ }\textbf {\bibinfo {volume} {02}},\ \bibinfo {pages} {030}
  (\bibinfo {year} {2023})},\ \Eprint {http://arxiv.org/abs/2210.15411}
  {arXiv:2210.15411 [hep-th]} \BibitemShut {NoStop}%
\bibitem [{\citenamefont {Feng}\ \emph {et~al.}(2011)\citenamefont {Feng},
  \citenamefont {Lust}, \citenamefont {Schlotterer}, \citenamefont
  {Stieberger},\ and\ \citenamefont {Taylor}}]{Feng:2010yx}%
  \BibitemOpen
  \bibfield  {author} {\bibinfo {author} {\bibfnamefont {W.-Z.}\ \bibnamefont
  {Feng}}, \bibinfo {author} {\bibfnamefont {D.}~\bibnamefont {Lust}}, \bibinfo
  {author} {\bibfnamefont {O.}~\bibnamefont {Schlotterer}}, \bibinfo {author}
  {\bibfnamefont {S.}~\bibnamefont {Stieberger}}, \ and\ \bibinfo {author}
  {\bibfnamefont {T.~R.}\ \bibnamefont {Taylor}},\ }\href {\doibase
  10.1016/j.nuclphysb.2010.10.013} {\bibfield  {journal} {\bibinfo  {journal}
  {Nucl. Phys. B}\ }\textbf {\bibinfo {volume} {843}},\ \bibinfo {pages} {570}
  (\bibinfo {year} {2011})},\ \Eprint {http://arxiv.org/abs/1007.5254}
  {arXiv:1007.5254 [hep-th]} \BibitemShut {NoStop}%
\bibitem [{\citenamefont {Caron-Huot}\ and\ \citenamefont
  {O'Connell}(2011)}]{Caron-Huot:2010nes}%
  \BibitemOpen
  \bibfield  {author} {\bibinfo {author} {\bibfnamefont {S.}~\bibnamefont
  {Caron-Huot}}\ and\ \bibinfo {author} {\bibfnamefont {D.}~\bibnamefont
  {O'Connell}},\ }\href {\doibase 10.1007/JHEP08(2011)014} {\bibfield
  {journal} {\bibinfo  {journal} {JHEP}\ }\textbf {\bibinfo {volume} {08}},\
  \bibinfo {pages} {014} (\bibinfo {year} {2011})},\ \Eprint
  {http://arxiv.org/abs/1010.5487} {arXiv:1010.5487 [hep-th]} \BibitemShut
  {NoStop}%
\bibitem [{\citenamefont {Novaes}\ and\ \citenamefont
  {Spehler}(1992)}]{Novaes:1991ft}%
  \BibitemOpen
  \bibfield  {author} {\bibinfo {author} {\bibfnamefont {S.~F.}\ \bibnamefont
  {Novaes}}\ and\ \bibinfo {author} {\bibfnamefont {D.}~\bibnamefont
  {Spehler}},\ }\href {\doibase 10.1016/0550-3213(92)90689-9} {\bibfield
  {journal} {\bibinfo  {journal} {Nucl. Phys. B}\ }\textbf {\bibinfo {volume}
  {371}},\ \bibinfo {pages} {618} (\bibinfo {year} {1992})}\BibitemShut
  {NoStop}%
\bibitem [{\citenamefont {Spehler}\ and\ \citenamefont
  {Novaes}(1991)}]{Spehler:1991yw}%
  \BibitemOpen
  \bibfield  {author} {\bibinfo {author} {\bibfnamefont {D.}~\bibnamefont
  {Spehler}}\ and\ \bibinfo {author} {\bibfnamefont {S.~F.}\ \bibnamefont
  {Novaes}},\ }\href {\doibase 10.1103/PhysRevD.44.3990} {\bibfield  {journal}
  {\bibinfo  {journal} {Phys. Rev. D}\ }\textbf {\bibinfo {volume} {44}},\
  \bibinfo {pages} {3990} (\bibinfo {year} {1991})}\BibitemShut {NoStop}%
\bibitem [{\citenamefont {Ochirov}(2018)}]{Ochirov:2018uyq}%
  \BibitemOpen
  \bibfield  {author} {\bibinfo {author} {\bibfnamefont {A.}~\bibnamefont
  {Ochirov}},\ }\href {\doibase 10.1007/JHEP04(2018)089} {\bibfield  {journal}
  {\bibinfo  {journal} {JHEP}\ }\textbf {\bibinfo {volume} {04}},\ \bibinfo
  {pages} {089} (\bibinfo {year} {2018})},\ \Eprint
  {http://arxiv.org/abs/1802.06730} {arXiv:1802.06730 [hep-ph]} \BibitemShut
  {NoStop}%
\bibitem [{\citenamefont {Chiodaroli}\ \emph {et~al.}(2022)\citenamefont
  {Chiodaroli}, \citenamefont {Johansson},\ and\ \citenamefont
  {Pichini}}]{Chiodaroli:2021eug}%
  \BibitemOpen
  \bibfield  {author} {\bibinfo {author} {\bibfnamefont {M.}~\bibnamefont
  {Chiodaroli}}, \bibinfo {author} {\bibfnamefont {H.}~\bibnamefont
  {Johansson}}, \ and\ \bibinfo {author} {\bibfnamefont {P.}~\bibnamefont
  {Pichini}},\ }\href {\doibase 10.1007/JHEP02(2022)156} {\bibfield  {journal}
  {\bibinfo  {journal} {JHEP}\ }\textbf {\bibinfo {volume} {02}},\ \bibinfo
  {pages} {156} (\bibinfo {year} {2022})},\ \Eprint
  {http://arxiv.org/abs/2107.14779} {arXiv:2107.14779 [hep-th]} \BibitemShut
  {NoStop}%
\bibitem [{\citenamefont {Dixon}(1996)}]{Dixon:1996wi}%
  \BibitemOpen
  \bibfield  {author} {\bibinfo {author} {\bibfnamefont {L.~J.}\ \bibnamefont
  {Dixon}},\ }in\ \href@noop {} {\emph {\bibinfo {booktitle} {{Theoretical
  Advanced Study Institute in Elementary Particle Physics (TASI 95): QCD and
  Beyond}}}}\ (\bibinfo {year} {1996})\ pp.\ \bibinfo {pages} {539--584},\
  \Eprint {http://arxiv.org/abs/hep-ph/9601359} {arXiv:hep-ph/9601359}
  \BibitemShut {NoStop}%
\bibitem [{\citenamefont {Hohm}\ \emph {et~al.}(2014)\citenamefont {Hohm},
  \citenamefont {Siegel},\ and\ \citenamefont {Zwiebach}}]{Hohm:2013jaa}%
  \BibitemOpen
  \bibfield  {author} {\bibinfo {author} {\bibfnamefont {O.}~\bibnamefont
  {Hohm}}, \bibinfo {author} {\bibfnamefont {W.}~\bibnamefont {Siegel}}, \ and\
  \bibinfo {author} {\bibfnamefont {B.}~\bibnamefont {Zwiebach}},\ }\href
  {\doibase 10.1007/JHEP02(2014)065} {\bibfield  {journal} {\bibinfo  {journal}
  {JHEP}\ }\textbf {\bibinfo {volume} {02}},\ \bibinfo {pages} {065} (\bibinfo
  {year} {2014})},\ \Eprint {http://arxiv.org/abs/1306.2970} {arXiv:1306.2970
  [hep-th]} \BibitemShut {NoStop}%
\bibitem [{\citenamefont {Huang}\ \emph
  {et~al.}(2016{\natexlab{b}})\citenamefont {Huang}, \citenamefont {Siegel},\
  and\ \citenamefont {Yuan}}]{Huang:2016bdd}%
  \BibitemOpen
  \bibfield  {author} {\bibinfo {author} {\bibfnamefont {Y.-t.}\ \bibnamefont
  {Huang}}, \bibinfo {author} {\bibfnamefont {W.}~\bibnamefont {Siegel}}, \
  and\ \bibinfo {author} {\bibfnamefont {E.~Y.}\ \bibnamefont {Yuan}},\ }\href
  {\doibase 10.1007/JHEP09(2016)101} {\bibfield  {journal} {\bibinfo  {journal}
  {JHEP}\ }\textbf {\bibinfo {volume} {09}},\ \bibinfo {pages} {101} (\bibinfo
  {year} {2016}{\natexlab{b}})},\ \Eprint {http://arxiv.org/abs/1603.02588}
  {arXiv:1603.02588 [hep-th]} \BibitemShut {NoStop}%
\bibitem [{\citenamefont
  {Lipinski~Jusinskas}(2019)}]{LipinskiJusinskas:2019cej}%
  \BibitemOpen
  \bibfield  {author} {\bibinfo {author} {\bibfnamefont {R.}~\bibnamefont
  {Lipinski~Jusinskas}},\ }\href {\doibase 10.1007/JHEP12(2019)143} {\bibfield
  {journal} {\bibinfo  {journal} {JHEP}\ }\textbf {\bibinfo {volume} {12}},\
  \bibinfo {pages} {143} (\bibinfo {year} {2019})},\ \Eprint
  {http://arxiv.org/abs/1909.04069} {arXiv:1909.04069 [hep-th]} \BibitemShut
  {NoStop}%
\bibitem [{\citenamefont {Koh}\ \emph {et~al.}(1987)\citenamefont {Koh},
  \citenamefont {Troost},\ and\ \citenamefont {Van~Proeyen}}]{Koh:1987hm}%
  \BibitemOpen
  \bibfield  {author} {\bibinfo {author} {\bibfnamefont {I.~G.}\ \bibnamefont
  {Koh}}, \bibinfo {author} {\bibfnamefont {W.}~\bibnamefont {Troost}}, \ and\
  \bibinfo {author} {\bibfnamefont {A.}~\bibnamefont {Van~Proeyen}},\ }\href
  {\doibase 10.1016/0550-3213(87)90642-0} {\bibfield  {journal} {\bibinfo
  {journal} {Nucl. Phys. B}\ }\textbf {\bibinfo {volume} {292}},\ \bibinfo
  {pages} {201} (\bibinfo {year} {1987})}\BibitemShut {NoStop}%
\bibitem [{\citenamefont {Britto}\ \emph
  {et~al.}(2005{\natexlab{b}})\citenamefont {Britto}, \citenamefont {Cachazo},\
  and\ \citenamefont {Feng}}]{Britto:2004ap}%
  \BibitemOpen
  \bibfield  {author} {\bibinfo {author} {\bibfnamefont {R.}~\bibnamefont
  {Britto}}, \bibinfo {author} {\bibfnamefont {F.}~\bibnamefont {Cachazo}}, \
  and\ \bibinfo {author} {\bibfnamefont {B.}~\bibnamefont {Feng}},\ }\href
  {\doibase 10.1016/j.nuclphysb.2005.02.030} {\bibfield  {journal} {\bibinfo
  {journal} {Nucl. Phys. B}\ }\textbf {\bibinfo {volume} {715}},\ \bibinfo
  {pages} {499} (\bibinfo {year} {2005}{\natexlab{b}})},\ \Eprint
  {http://arxiv.org/abs/hep-th/0412308} {arXiv:hep-th/0412308} \BibitemShut
  {NoStop}%
\bibitem [{\citenamefont {Bianchi}\ \emph {et~al.}(2011)\citenamefont
  {Bianchi}, \citenamefont {Lopez},\ and\ \citenamefont
  {Richter}}]{Bianchi:2010es}%
  \BibitemOpen
  \bibfield  {author} {\bibinfo {author} {\bibfnamefont {M.}~\bibnamefont
  {Bianchi}}, \bibinfo {author} {\bibfnamefont {L.}~\bibnamefont {Lopez}}, \
  and\ \bibinfo {author} {\bibfnamefont {R.}~\bibnamefont {Richter}},\ }\href
  {\doibase 10.1007/JHEP03(2011)051} {\bibfield  {journal} {\bibinfo  {journal}
  {JHEP}\ }\textbf {\bibinfo {volume} {03}},\ \bibinfo {pages} {051} (\bibinfo
  {year} {2011})},\ \Eprint {http://arxiv.org/abs/1010.1177} {arXiv:1010.1177
  [hep-th]} \BibitemShut {NoStop}%
\bibitem [{\citenamefont {Feng}\ and\ \citenamefont
  {Taylor}(2012)}]{Feng:2011qc}%
  \BibitemOpen
  \bibfield  {author} {\bibinfo {author} {\bibfnamefont {W.-Z.}\ \bibnamefont
  {Feng}}\ and\ \bibinfo {author} {\bibfnamefont {T.~R.}\ \bibnamefont
  {Taylor}},\ }\href {\doibase 10.1016/j.nuclphysb.2011.11.004} {\bibfield
  {journal} {\bibinfo  {journal} {Nucl. Phys. B}\ }\textbf {\bibinfo {volume}
  {856}},\ \bibinfo {pages} {247} (\bibinfo {year} {2012})},\ \Eprint
  {http://arxiv.org/abs/1110.1087} {arXiv:1110.1087 [hep-th]} \BibitemShut
  {NoStop}%
\bibitem [{\citenamefont {Guevara}\ \emph {et~al.}(2019)\citenamefont
  {Guevara}, \citenamefont {Ochirov},\ and\ \citenamefont
  {Vines}}]{Guevara:2018wpp}%
  \BibitemOpen
  \bibfield  {author} {\bibinfo {author} {\bibfnamefont {A.}~\bibnamefont
  {Guevara}}, \bibinfo {author} {\bibfnamefont {A.}~\bibnamefont {Ochirov}}, \
  and\ \bibinfo {author} {\bibfnamefont {J.}~\bibnamefont {Vines}},\ }\href
  {\doibase 10.1007/JHEP09(2019)056} {\bibfield  {journal} {\bibinfo  {journal}
  {JHEP}\ }\textbf {\bibinfo {volume} {09}},\ \bibinfo {pages} {056} (\bibinfo
  {year} {2019})},\ \Eprint {http://arxiv.org/abs/1812.06895} {arXiv:1812.06895
  [hep-th]} \BibitemShut {NoStop}%
\bibitem [{\citenamefont {Ballav}\ and\ \citenamefont
  {Manna}(2021)}]{Ballav:2020ese}%
  \BibitemOpen
  \bibfield  {author} {\bibinfo {author} {\bibfnamefont {S.}~\bibnamefont
  {Ballav}}\ and\ \bibinfo {author} {\bibfnamefont {A.}~\bibnamefont {Manna}},\
  }\href {\doibase 10.1007/JHEP03(2021)295} {\bibfield  {journal} {\bibinfo
  {journal} {JHEP}\ }\textbf {\bibinfo {volume} {03}},\ \bibinfo {pages} {295}
  (\bibinfo {year} {2021})},\ \Eprint {http://arxiv.org/abs/2010.14139}
  {arXiv:2010.14139 [hep-th]} \BibitemShut {NoStop}%
\end{thebibliography}%
\bibliographystyle{apsrev4-1}

\end{document}